\documentclass[journal]{IEEEtran}
\IEEEoverridecommandlockouts
\usepackage{cite}
\usepackage{amsmath,amssymb,amsfonts}
\usepackage{algorithmic}
\usepackage{graphicx}
\usepackage{textcomp}
\usepackage{url} 
\usepackage{longtable} 
\usepackage{booktabs}
\usepackage[utf8]{inputenc} 
\usepackage[T1]{fontenc} 
\usepackage{algorithm}
\usepackage{comment}
\usepackage{multirow} 
\usepackage{makecell}

\def\BibTeX{{\rm B\kern-.05em{\sc i\kern-.025em b}\kern-.08em
    T\kern-.1667em\lower.7ex\hbox{E}\kern-.125emX}}

\title{Speech as a Biomarker for Disease Detection}
\author{\IEEEauthorblockN{Catarina Botelho\textsuperscript{\dag},  Alberto Abad\textsuperscript{\dag}, Tanja Schultz\textsuperscript{\ddag} and Isabel Trancoso\textsuperscript{\dag}}\\ 
\IEEEauthorblockA{\textsuperscript{\dag}INESC-ID \&
Instituto Superior T\'{e}cnico, University of Lisbon, Portugal\\
\textsuperscript{\ddag}Cognitive Systems Lab (CSL), University of Bremen, Germany\\
\small\texttt{\{catarina.t.botelho,alberto.abad,isabel.trancoso\}@inesc-id.pt,tanja.schultz@uni-bremen.de}}}

\begin{document}

\maketitle

\begin{abstract}
Speech is a rich biomarker that encodes substantial information about the health of a speaker, and thus it has been proposed for the detection of numerous diseases, achieving promising results. However, questions remain about what the models trained for the automatic detection of these diseases are actually learning and the basis for their predictions, which can significantly impact patients' lives.
This work advocates for an interpretable health model, suitable for detecting several diseases, motivated by the observation that speech-affecting disorders often have overlapping effects on speech signals. 
A framework is presented that first defines "reference speech" and then leverages this definition for disease detection. Reference speech is characterized through reference intervals, i.e., the typical values of clinically meaningful acoustic and linguistic features derived from a reference population. This novel approach in the field of speech as a biomarker is inspired by the use of reference intervals in clinical laboratory science.
Deviations of new speakers from this reference model are quantified and used as input to detect Alzheimer's and Parkinson's disease. The classification strategy explored is based on Neural Additive Models, a type of glass-box neural network, which enables interpretability.
The proposed framework for reference speech characterization and disease detection is designed to support the medical community by providing clinically meaningful explanations that can serve as a valuable second opinion.

\end{abstract}


\section{Introduction}
\label{sec:introduction}

{O}{verburdened} 
health systems worldwide face numerous challenges, exacerbated by an aging population.
Speech, a rich and ubiquitous biomarker, allied with highly accurate machine learning systems, offers the potential for low-cost, widespread detection of several diseases. 
This potential stems from the involvement of the respiratory, nervous, and muscular systems in speech production. This implies that disruptions in any of these systems can perturb the speech signal. Consequently, speech can encode information indicative of diseases affecting these systems, going beyond the so-called speech and language disorders (e.g. sigmatism, stuttering), and including\footnote{The following categories do not constitute a formal classification. For an official categorization, refer to the International Classification of Diseases, 11th Revision (ICD-11)\cite{icd11}.} neurodegenerative diseases, such as Parkinson's Disease (PD)~\cite{pompili2017PKDetection, correia2021wsmICASSP, Parkinson-x-vectors}, Alzheimer's Disease (AD)~\cite{Alzheimer-x-vectors, weiner2016speechbased, lopez2018Alzheimer, ablimit_botelho2022exploring}, and Multiple Sclerosis~\cite{noffs2018speech}; psychiatric disorders such as depression \cite{correia2021wsmICASSP, afshan2018effectiveness}, and schizophrenia~\cite{parola2020voice}; and diseases that concern respiratory organs, such as Obstructive Sleep Apnea (OSA) \cite{botelho2019osa, perero2019adversOsa}, and COVID-19~\cite{deshpande2020overview_covid}.
Beyond the mentioned examples, there is a vast literature on automatic classification systems leveraging speech to perform the binary classification of healthy controls and each of these diseases, reporting very promising results.

In real scenarios however, these diseases often co-exist. The coexistence of two or more chronic conditions in the same individual, or \textit{multimorbidity},
has been rising in prevalence over recent years \cite{who_2016_multimorbidity, barnett2012epidemiology_multimorbidity, wang2014epidemiology_multimorbidity}. 
The World Health Organization emphasizes that healthcare of people with multiple conditions should be provided by medical generalists who combine a community base and comprehensive clinical skills with “interpretive medicine”, integrating multiple sources of knowledge with individual needs assessment \cite{who_2016_multimorbidity}. 
Likewise, we hypothesize that a speech-based tool to support medical diagnosis and monitoring of chronic conditions should adopt a holistic approach, facilitating the interpretative assessment of multiple diseases. This is relevant, as some diseases are risk factors for others, and their effects on speech signals can overlap. 
Moreover, the ongoing diversification of medical disciplines increases the difficulty of identifying all diseases for a single specialist. This further underscores the need for a comprehensive, speech-based diagnostic tool that can assist in the identification and monitoring of multiple conditions.

However, existing datasets for disease detection are often small and labeled only for individual diseases. The naive combination of different datasets containing recordings from individuals with a single specific disease to perform a cross-corpora study for multi-disease classification can result in unreliable results that would not generalize to unseen recording conditions~\cite{botelho2022healthy_speech, ablimit_botelho2022exploring}. 
It has also been established that small datasets may lead to overoptimistic estimation of performance, with models learning confounding variables and overfitting to the dataset~\cite{berisha2022reported}. 
In light of the aforementioned considerations, we claim that a valuable step towards the adoption of speech and language technologies in real health applications would be to obtain a definition of reference speech that could be used independently of the dataset of origin, and subsequently be applied for the identification of disease signatures.

In this study, \textit{reference speech} refers to the speech characteristics common to a reference population, ideally comprising healthy individuals of varying ages and biological sexes. Acknowledging the challenges of defining health and the prevalence of subclinical disease\footnote{Unlike a clinical disease which has identifiable signs and symptoms, a subclinical disease lacks recognizable clinical findings. Many diseases (e.g. diabetes) often remain subclinical before manifesting clinically~\cite{subclinical_def}.}, 
we do not assert that our reference population consists exclusively of healthy speakers. Instead, we utilize the speech of individuals who self-report as disease-free.

We propose to characterize reference speech through \textit{reference intervals} (RIs) of clinically meaningful speech and language features. RIs represent the typical range of values for specific parameters within a reference population. In this context, RIs are computed as the 2.5th and 97.5th percentiles of the distribution of each parameter within the reference population. Ideally, the speech characteristics of an unseen healthy individual should fall within the RIs derived from the reference population.
The concept of RIs is commonly applied in clinical laboratory science to interpret laboratory results and assess individual health. 
The idea of characterizing reference speech using reference intervals was first introduced in our previous work~\cite{botelho2023RIs}. 

The definition of reference speech is subsequently used to perform the detection of speech affecting diseases. 
Each disease detection task is formulated as a binary classification problem (patients versus controls) and addressed using Neural Additive Models (NAMs)~\cite{agarwal2021nams}. These are interpretable neural networks that provide an insight into the decision process, which can be of utmost importance in the medical domain, especially to avoid the models learning spurious correlations.

While the proposed framework is designed to be suitable for different speech affecting diseases, we showcase it for AD and PD due to the availability of public datasets annotated for each of these diseases. 
This work presents a unified framework with distinct models for disease detection. In the future, this framework could enable the simultaneous detection of multiple diseases, provided that datasets with similar tasks collected under comparable conditions are available.

The primary contributions of this research are: (1) an overview of the most notorious effects on speech commonly associated with frequently studied speech-affecting diseases, along with their mediating mechanisms, highlighting the importance of a holistic perspective of speech as a biomarker for health; (2) a framework for defining reference speech building on our prior work~\cite{botelho2023ref_speech}; and (3) the application of NAMs for the detection of speech-affecting diseases, leveraging the reference speech definition and providing interpretable decisions. These contributions are foundations to our overall vision that goes beyond the experiments described in this work, and advocates for the usage of speech as a biomarker for health monitoring in general, rather than focusing on single diseases. 

The rest of this document is organized as follows. Section~\ref{sec:challenges} discusses challenges in the current state of the art on the detection of speech affecting diseases, particularly their partially overlapping manifestations on speech, and the overoptimistic performance estimates in the literature. Section~\ref{sec:related_work-RIs} presents related work on characterizing healthy speech, and on reference interval estimation. 
Section~\ref{sec:framework-and-corpora} introduces the corpus that constitutes the reference population for RI estimation, and the corpora used for disease detection.
Section~\ref{sec:task1-ref-speech} dives into the first stage of this work, the definition of reference speech, describing the methodology and presenting the results. Section~\ref{sec:task2-disease-detection} presents the methodology and the results for the second stage of this work, where the definition of reference speech is leveraged for the detection of AD and PD.
Section~\ref{sec:limitations} discusses the limitations of our approach. Finally, section~\ref{sec:conclusion} presents the main conclusions. Further details are provided in the appendices and in~\cite{PhDthesis2024Botelho}.

%%%%%%%%%%%%%%%%%%%%%%%%%%%%%%%%%%%%%%%%%%%%%%%%%%%%%%%%%%%%%%%%%%%%%%%%%%%%%%%%%%%%%%%%%%%%%%%%%%%%%%%%%%%%%%%%%%%%%%%%%%%%%%%%%%%%%%%%%%
\section{Challenges in the automatic detection of speech affecting diseases}
\label{sec:challenges}

\begin{figure*}[t]
  \centering
  \includegraphics[width=\textwidth]{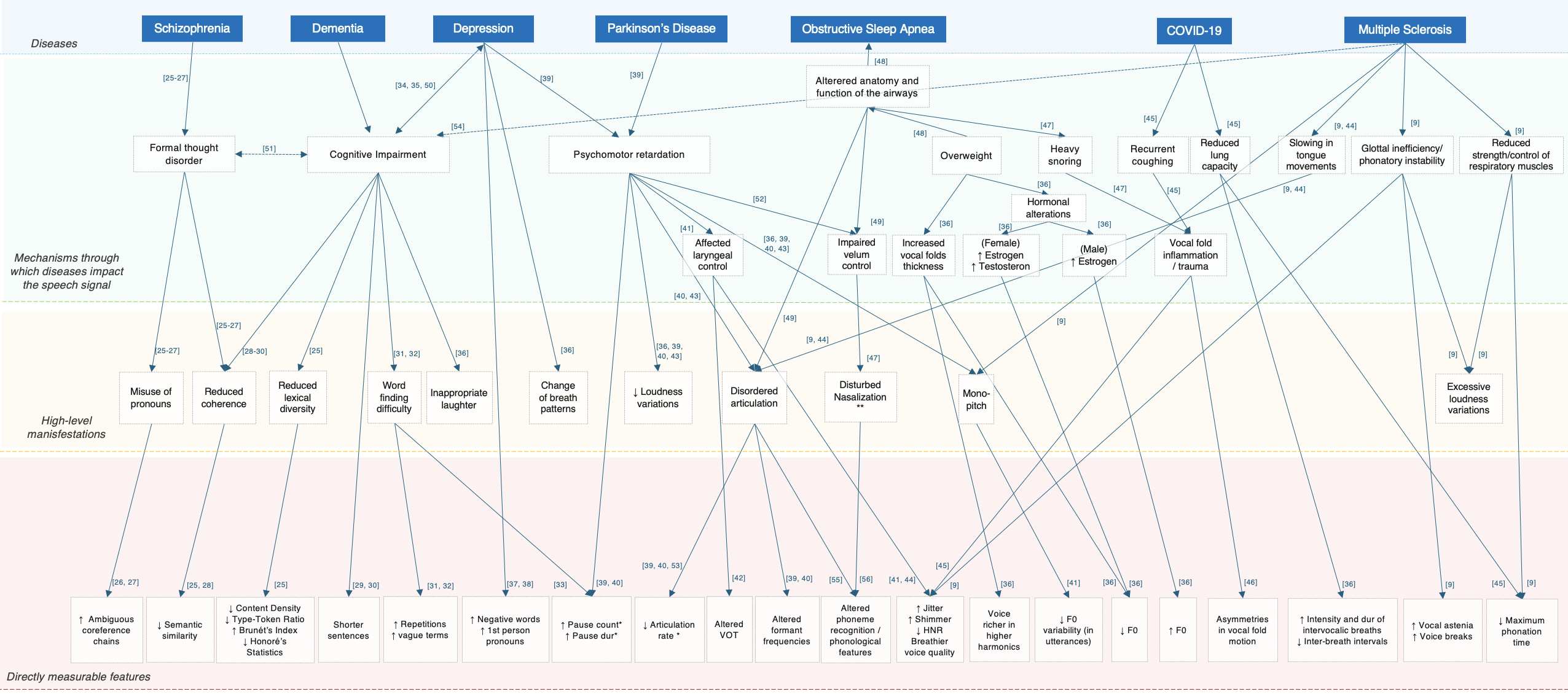}
  \caption{Examples of mechanisms through which speech affecting diseases impact the speech signal.  
  The diagram includes references \cite{voleti_berisha2019review, bedi2015psychosis, iter2018schizophrenia, sanz2022coherenceAD, pompili2019phd, hier1985language_dementia, forbes2002distinct, oppenheim1994earliest_ad, hoffmann2010temporal_ad, krishnan2002comorbidity, laird2019late-life-depression-psychobiological, ritasingh2019profiling, tolboll2019depression_1stppronouns, rude2004language_depression, flint1993abnormal_depression, cummins2015review, ma2020voicePD, vasquez2017convolutional, ramig2008speech_pd, hecker2022voice, asiaee2020voice_covid, al2021covid_vocal_folds, OSA6-pozo2009, malhotra2002osa_lancet, monoson_and_fox1987preliminary, halahakoon2019cognitive, kerns2002cognitive, hoodin1989nasal, martinez2016speech, noffs2021speech, noffs2018speech, jiao_berisha2017interpretable, saxon2019objective}. (*)~articulation rate and pauses have been reported to be associated with depression, via psychomotor retardation, however, although psychomotor retardation is also present in PD, these features have inconsistent reports for PD~\cite{skodda2011aspects}. (**)~Disturbed nasalization, as a consequence of impaired velum control, is associated with PD via psychomotor retardation and with OSA. In PD an increase in nasal airflow is reported~\cite{hoodin1989nasal}, while for OSA, a smaller difference between nasal and oral sounds is reported~\cite{monoson_and_fox1987preliminary}. 
  }
\label{fig:multi_diseases_impact_speech}
\vspace{-12pt}
\end{figure*}

\subsection{\textbf{Overlapping manifestations in speech and multimorbidity}}

The diagram in Fig.~\ref{fig:multi_diseases_impact_speech} illustrates several speech affecting diseases, the mechanisms through which they impact the speech signal, and examples of speech features that capture such alterations. This figure is not exhaustive but aims to provide an overview of the most notable speech effects associated with these diseases.
Some of the references depicted in this figure provide very complete overviews on the use of speech analysis for the detection of some of these diseases.
For example, Boschi et al.~\cite{boschi2017connected} provided an exhaustive review on the use of spontaneous speech tasks to characterize language disorders in prevalent neurodegenerative diseases. 
Hecker et al.~\cite{hecker2022voice} systematically reviewed voice analysis for recognizing neurological and psychiatric diseases. 
Voleti et al.~\cite{voleti_berisha2019review} reviewed speech and language features for cognitive and thought disorder analysis. 
Cummins et al.~\cite{cummins2015analysis} described the impact of depression and suicidality on paralinguistic speech characteristics and their application in classification systems. 
Ma et al.~\cite{ma2020voicePD} discussed voice changes in PD patients, linking these changes to physiological and anatomical characteristics. 
Deshpande and Schuller~\cite{deshpande2020covid_review} provided an overview of COVID-19 screening and monitoring through speech and other human generated audio signals, also reviewing relevant research on respiratory conditions such as asthma and OSA.

However, to the best of our knowledge, no existing works systematize the overlapping effect of multiple diseases in speech together with the main mediating mechanisms. 
Our diagram represents an effort in this direction, aiming to enhance the understanding of how speech could serve as a biomarker for multidisease detection. 
Singh~\cite{ritasingh2019profiling} defends that causal relationships between parameters and voice must be sought, or reasonably guessed, and then features should be selected to capture these causal relationships. This diagram facilitates such reasoning. 
For instance, if COVID-19 is hypothesized to cause vocal fold inflammation due to repetitive coughing, features that capture such inflammation should be derived to predict COVID-19.
Furthermore, when studying a different disease not included in the diagram, one can consider whether it shares any listed mechanisms, such as association with overweight or reduced lung capacity, to anticipate potential speech alterations.
It should be emphasized that most mechanisms illustrated are hypotheses presented in the literature and may not be present in all cases. Further research and multidisciplinary discussion are needed to validate these hypotheses.

In the diagram, it is notorious that the impact of certain diseases on the speech signal may overlap with that of other diseases. For example, both depression and PD are associated with psychomotor retardation, and thus have similar manifestations on the speech signal. 
It also becomes clear that often these speech features that capture speech alterations are non specific for a single disease, and thus, when considered alone, may be insufficient for the automatic detection of diseases. F0 based features, for instance, appear to be altered as a result of several diseases, not to mention possible alterations associated with healthy aging~\cite{ritasingh2019profiling}, or even healthy changes across the menstrual cycle in fertile women~\cite{bryant2009vocalCuesOvulation}. 
It is also important to refer that some of the features depicted in the diagram have had inconsistent reports in the literature. For example, voicing onset time (VOT) has been found both higher and lower in people suffering from PD when compared with healthy controls \cite{fischer2010vot}. The frequencies of formants have also been inconsistently reported to change with depression \cite{cummins2015review}. 
Furthermore, there are many other factors that directly or indirectly impact the speech production process, and thus introduce alterations on the speech signal. Some of these factors include medication~\cite{ma2020voicePD, goberman2002phonatory, pompili2020assessment}, or other medical interventions, emotions or mental states.

Besides the fact that these diseases have overlapping effects on the speech signal, it is important to note that they are often considered risk factors for each other, and thus are likely to co-exist. 
PD and AD, for instance, are risk factors for depression, and the converse is also true~\cite{krishnan2002comorbidity, greenwald1995depression}.
OSA is also associated with depression, potentially mediated by disturbed sleep patterns and obesity, a major risk factor for OSA. Additionally, OSA is linked to cognitive impairments, possibly due to repetitive hypoxemia~\cite{vanek2020osa_depression_dementia}.

%%%%%%%%%%%%%%%%%%%%%%%%%%%%%%%%%%%%%%%%%%%%%%%%%%%%%%%%%%%%%%%%%%%%%%%%%%
\subsection{\textbf{Data scarcity}}

Despite promising results, the literature on disease detection from voice and language analysis often reports overoptimistic findings.
Berisha et al. \cite{berisha2022reported} found that smaller sample sizes are associated with inflated accuracy in dementia detection from speech, suggesting publication bias and overfitting. 
In a recent survey talk\footnote{N. Cummins. Machine Learning for Speech-based Health Analysis: State-of-the-art and Future Challenges. Survey talk at Interspeech 2022 \url{https://www.interspeech2022.org/program/surveytalk.php}}, Cummins has also referred to this problem and cautioned against the Clever Hans Effect in this context.
Ozbolt et al.~\cite{ozbolt2022things} identified several methodological issues that could lead to overoptimistic results in classifying PD patients and healthy controls using sustained vowels, including age-unmatched classes, large feature vectors, and data leakage between train and test sets. 
Espinoza et al.~\cite{OSA5-espinoza2016} highlighted two main limitations in OSA detection from speech analysis:  the influence of confounding factors (e.g., age, height, sex), and overfitting of feature selection and validation methods when working with a high dimensional feature set compared to the number of samples.

The frequently reported overoptimistic results due to confounding factors and overfitting on small datasets, call for more trustworthy research on the use of speech as a biomarker for disease detection. Namely, using interpretable models may represent a step towards assuring that the models are learning properties indeed attributable to the disease, and not confounding factors.

Our previous work~\cite{solera2021covid} observed the consequences of unexpected confounding factors, specifically how the bandwidth of the speech signal affected the ComParE 2021 challenge corpus for COVID-19 detection \cite{Schuller2021compare_covid}, questioning the validity of black-box classifiers.
Another study \cite{botelho2022healthy_speech} found that standard features for disease detection contain substantial information about recording conditions, enabling both supervised and unsupervised detection of the source dataset.

\section{Related work}
\label{sec:related_work-RIs}

\subsection{\textbf{Characterizing reference speech}}
Some researchers have delved into characterizing features in the context of healthy speech, providing means and standard deviations.
Teixeira and Fernandes~\cite{teixeira2014jitter} studied jitter, shimmer, and harmonics-to-noise ratio (HNR) in 34 female and 7 male speakers, focusing on sustained vowels /a/, /i/, and /u/ at various tones. No other acoustic features were included in this study.
Shivkumar et al.~\cite{shivkumar2020blabla_feats} proposed a toolkit for extracting clinically meaningful linguistic features, and present statistics for these features for the healthy speakers in the AMI meeting corpus \cite{carletta2005ami_corpus}. 
This study, however, did not include any acoustic feature, since its purpose was to illustrate the toolkit.
Hence, the aim of these two studies was not to define or comprehensively characterize reference speech. 

Schwoebel et al.~\cite{schwoebel2021voiceome}, on the other hand, introduced the Voiceome Protocol and the corresponding Voiceome Dataset as standards to characterize healthy speech. The authors reported means and standard deviations for several acoustic and linguistic features, broken down by age range and gender, on their GitHub\footnote{\url{https://github.com/jim-schwoebel/voiceome}}, though the dataset is not publicly available.

\subsection{\textbf{Reference intervals in clinical laboratory science}}

Reference intervals (RIs) are crucial in clinical laboratory science for interpreting quantitative pathology results, such as those from hematology tests~\cite{jones2018indirect}. RIs, defined by a lower and upper reference limit, represent the expected range of values in a reference population~\cite{ozarda2018distinguishing}. Laboratory results outside the RI do not necessarily imply disease but indicate the need for further medical evaluation~\cite{ozarda2018distinguishing}.

The traditional, or \textit{direct approach}, for determining RIs involves \textit{selecting} a reference population (minimum of 120 individuals per partition, e.g. sex, age range) respecting pre-defined criteria, \textit{collecting samples} for that population, performing statistical evaluation using non parametric methods and outlier removal, and estimating the RI between the two reference limits~\cite{horowitz2010ep28a3c}.
This method faces challenges such as defining health, the presence of subclinical disease, and selection bias associated with small cohorts~\cite{jones2018indirect}. Further guidelines can be found in~\cite{horowitz2010ep28a3c}.

An alternative approach, known as the \textit{indirect approach}, mines data from existing pathology databases, i.e., it is \textit{based on laboratory results collected for other purposes}, usually for routine clinical care. These databases include results from diseased patients but also from healthy subjects, allowing for the extraction of the underlying reference distributions~\cite{ozarda2018distinguishing}. This approach is faster, cheaper, avoids patient inconvenience, and circumvents ethical issues related to informed consent from vulnerable populations\cite{jones2018indirect, ozarda2016reference}, while providing extensive data for analysis. 
However, the presence of diseased sub-populations can influence RIs, and at least 400 subjects per partition are recommended \cite{jones2018indirect}.

While guidelines favor the direct approach \cite{horowitz2010ep28a3c}, the indirect approach is increasingly popular, especially in pediatric and geriatric populations where data sampling is more challenging.

If the underlying distribution of the data in the reference population is Gaussian, the reference limits that constitute the RI corresponds to the $mean \pm 1.96 \times std$, in which $std$ stands for standard deviation~\cite{ichihara2010appraisal}. If such assumption cannot be made, which is frequently the case for the studies of RI estimation, either data must be first transformed to a Gaussian distribution, e.g. using a Box-Cox power transformation, or a non-parametric estimation can be made. In the case of a non-parametric estimation, the limits of the RI correspond to the $2.5^{th}$ and $97.5^{th}$ percentile~\cite{ichihara2010appraisal}. 
Both the non-parametric approach and the power transformation of data followed by the parametric approach provide similar results according to \cite{ichihara2010appraisal}, but the non-parametric method is recommended in some studies
\cite{horowitz2010ep28a3c}.

%%%%%%%%%%%%%%%%%%%%%%%%%%%%%%%%%%%%%%%%%%%%%%%%%%%%
\section{Corpora}
\label{sec:framework-and-corpora}

\begin{figure*}[t!]
  \centering
\includegraphics[width=0.97\linewidth]{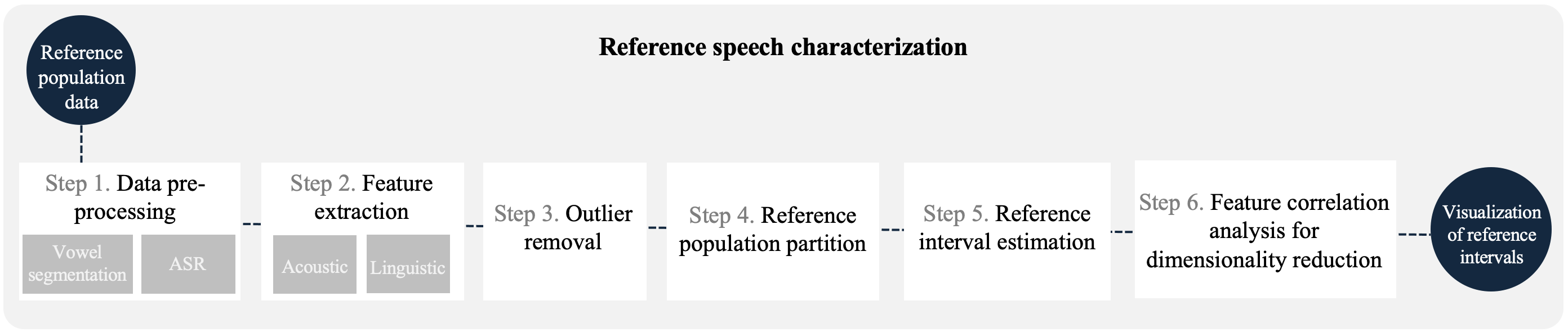}
\vspace{-4pt}
  \caption{Overview of the steps entailed for reference speech characterization.}
  \label{fig:pipeline_task_1}
  \vspace{-12pt}
\end{figure*}

The analysis conducted in this work can be subdivided into two main stages: first, the definition of reference speech, and second, leveraging this definition to perform the detection of speech affecting diseases.
Each stage demands for specific data. Below, we introduce the corpora used for this study.

\subsection{\textbf{Reference Population Data}}

Aligned with the indirect approach for estimating RIs described in Section~\ref{sec:related_work-RIs}, in this study, the reference population used to derive RIs was sourced from already existing databases. Since routine pathology databases typically lack speech recordings, we used the  the Crowdsourced Language Assessment
Corpus (CLAC)~\cite{haulcy21_interspeech}, a corpus spoken in English, created to provide a collection of audio samples from healthy speakers. 
CLAC includes speech from 1,832 speakers, claiming to have no health-related symptoms that might affect their speech. 
Besides presumably having a low incidence of unhealthy subjects, and including standard speech tasks for disease detection, CLAC offers the advantage of being larger than other publicly available speech corpora in the field that studies speech as a biomarker.

The subsets of the corpus used in this study include two picture description tasks  (Cookie Theft picture and Picnic picture), and a sustained vowel task (/a/). The 13 speakers that identify as "other" in terms of gender were excluded from the analysis because the sample size was too small. Information on the gender and age of the included speakers is provided later on, in Table~\ref{tab:ref_population}.

Although integrating multiple datasets would be ideal, this study focuses on picture description and sustained vowel tasks to enable comparison with publicly available datasets for Alzheimer's disease and Parkinson's disease detection. To the best of our knowledge, no other publicly available datasets include these tasks with healthy speakers, or speakers claiming to be disease-free.

\subsection{\textbf{Datasets for disease detection}}
Two additional datasets were used to investigate disease detection and to conduct a comparative analysis with the reference speech. The datasets employed were ADReSS~\cite{ADReSS} and PC-GITA~\cite{orozco2014new}, for the analysis of Alzheimer's and Parkinson's disease, respectively.

ADReSS comprises speech recordings of 156 subjects describing the Cookie Theft picture, 78 controls and 78 AD patients, matched for age and gender. This dataset is a subset of the Pitt corpus \cite{dementia_bank} curated for the 2020 ADReSS Challenge~\cite{ADReSS}.
Although the audios released in the ADReSS challenge were acoustically enhanced, the experiments conducted used the original version made available in the Pitt Corpus~\cite{dementia_bank}. The interviewers' interventions were removed, using the annotations provided in the manual transcriptions.

The Parkinson’s disease corpus from the Applied Telecommunications Group (PC-GITA) includes recordings of 100 subjects, 50 PD patients, and 50 controls matched by age and gender. The corpus is spoken in Colombian Spanish, and the recordings were captured in noise controlled conditions. To allow a fair comparison with the reference intervals determined for an English speaking reference population, the only task explored in this work is the enunciation of a sustained vowel /a/ (three repetitions per subject).
PC-GITA was made available for the 2015 ComParE challenge~\cite{schuller2015interspeech}, for the assessment of PD severity.

%%%%%%%%%%%%%%%%%%%%%%%%%%%%%%%%%%%%%%%%%%%%%%%%%%%%%%
\section{Reference speech characterization (RSC)}
\label{sec:task1-ref-speech}

This section describes our proposed approach to characterize reference speech by defining reference intervals for a set of literature-informed, knowledge-based features. This method is inspired by the indirect approach for RI estimation, described in section~\ref{sec:related_work-RIs}.

\subsection{\textbf{Reference Speech Characterization Pipeline}}
\label{sec:task1-ref-speech-method}

The definition of reference speech started by a pre-processing step which included Automatic Speech Recognition (ASR) or vowel segmentation whenever appropriate (step 1). This step was followed by the extraction of interpretable acoustic and linguistic features (step 2), and by the removal of outliers (step 3). Subsequently, we assessed the necessity of partitioning the reference population based on gender, age ranges, or speech tasks (step 4). The RIs for each feature were then established using the refined reference population (step 5). Additionally, a dimensionality reduction strategy was investigated to enhance interpretability (step 6). Fig.~\ref{fig:pipeline_task_1} presents an overview of these steps.
The method follows our previous work~\cite{botelho2023ref_speech}, with some enhancements in each step, and with the addition of step 6.

\vspace{10pt}
\noindent \textbf{Step 1. Data pre-processing}
\label{sec:c6-method-vowel-processing}

\noindent
The data pre-processing step involves the application of different sub-steps depending on the nature of the speech task. If the task involves a sustained vowel, vowel segmentation is employed. Conversely, for spontaneous speech tasks, ASR is applied.

\paragraph*{Vowel segmentation}

Due to its crowdsourced nature, CLAC includes recordings with anomalies, particularly in sustained vowels (e.g., low energy, gain decrease due to unrecognized speech). 
To improve the overall quality of the recordings that constitute the reference population,  data filtering was performed to remove or segment sustained vowel recordings from CLAC, as detailed in Appendix~\ref{sec:appendix_data_preprocess_feature_ext}.

\paragraph*{Automatic speech recognition} 

The extraction of linguistic features required transcriptions of the speech recordings. Unlike previous studies optimizing ASR systems for individuals with Parkinson's or Alzheimer's disease (e.g., \cite{hu2023exploring}), this study adopted a zero-shot approach suitable for the general population and various speech-affecting diseases.  
Our previous analysis in~\cite{botelho2024ad_llms}, which compared five state-of-the-art ASR systems 
(\textit{wavlm-libri-clean-100h-large}~\cite{Chen2021WavLMLS},
\textit{wav2vec2-large-960h}~\cite{10.5555/3495724.3496768},
\textit{wav2vec2-large-robust-ft-swbd-300h}~\cite{hsu21_interspeech},
\textit{wav2vec2-large-xlsr-53-english}~\cite{grosman2021xlsr53-large-english},
\textit{whisper-large}~\cite{whisper-large}),
concluded that \textit{whisper-large} (henceforth referred to as whisper) achieves the lowest word error rate (26.9\%) on ADReSS, likely due to its training on 680,000 hours of supervised data from the web~\cite{whisper-large}. Therefore, our analysis was conducted based on whisper transcriptions.
However, whisper often outputs transcriptions cleaner than the actual audio in terms of fluency, namely by removing fillers or repetitions, which may encode relevant information for studying cognitive impairment. Therefore, additional experiments were conducted on the second-best model, \textit{wav2vec2-large-robust-ft-swbd-300h}, which retains such disfluencies but sometimes produces non-existent words, potentially affecting downstream tasks. The results of these experiments are reported in the appendices.

It is noteworthy that whisper's training data has not been disclosed, raising the possibility that datasets such as ADReSS and CLAC may have been seen during training.

\vspace{10pt}
\noindent \textbf{Step 2. Feature Extraction}
\label{sec:c6-method-features}

\begin{table*}[ht!]
\caption{Description of the features used. Observations: 
In rhythm-related features, when the descriptions refer to the total time, it assumes that silences before the start and after the end of the speech signal were removed, unless explained otherwise. 
TTR stands for type-to-token ratio, and HNR to harmonics-to-noise ratio.
} 
\label{tab:features_description}  
\setlength{\tabcolsep}{4 pt}
\resizebox{\linewidth}{!}{
\scriptsize
\begin{tabular}{p{1cm} p{2cm} p{2cm} p{1.1cm} p{9cm}}

\toprule
\textbf{Category} &  \textbf{Feature Name} & \textbf{Functional} & \textbf{Method} & \textbf{Description}  \\
\hline
 &  Content density & -- & BlaBla & Proportion of number of open class words, i.e. nouns, verbs, adjectives and adverbs,  to the number of close class words, i.e. determiners, pronouns, conjunctions and prepositions~\cite{shivkumar2020blabla_feats}. \\
 &  Idea density &  -- & BlaBla & Proportion of verbs, adjectives, adverbs, prepositions and conjunctions to all words across sentences~\cite{shivkumar2020blabla_feats}. \\
 &  Honoré statistic &  -- & BlaBla & Calculated as $(100*log(N))/(1-(V1)/(V))$, where $V$ is number of unique words, $V1$ is the number of words in the vocabulary only spoken once, and $N$ is overall text length~\cite{shivkumar2020blabla_feats}. \\
 &  \raggedright Brunet's Index &  -- & BlaBla & Calculated as $N^(V^{-0.165})$, where $V$ is number of unique words and $N$ is overall text length. Measures the lexical richness. It is a version of TTR, insensitive to text-length~\cite{shivkumar2020blabla_feats}. \\
 &  TTR &  -- & BlaBla & The number of word types divided by the number of word tokens~\cite{shivkumar2020blabla_feats} . \\
 & \raggedright Discourse marker rate &  -- & BlaBla & The rate of discourse markers across all sentences~\cite{shivkumar2020blabla_feats}  (eg. "so, ok, anyway, right"~\cite{discourseMarkers-cambridgeDic}). \\
 &  Polarity &  -- & TextBlob & Varies between $[-1,1]$, where $-1$ defines a negative sentiment and $1$ defines a positive sentiment. \\
\centering \textbf{Content} &  Repetition ratio &  -- & dedicated script & Number of repeated words over total number of words \\
 & \raggedright First person pronouns &  -- & dedicated script & Ratio of number of personal pronouns ("i", "me", "mine", "my"), to the text length. \\
 
 &  Coherence &  \raggedright mean, variability & cosine similarity & Cosine similarity between sentence embeddings of adjacent text segments (14 tokens), computed with the pretrained sentence-transformer model \textit{all-mpnet-base-v2}. (More details in the text.) \\
 &  \raggedright Coreference chain ratio &  -- & wl-coref & Number of coreference chains over text length. \\
 &  \raggedright Ambiguous coreference chain &  -- & wl-coref & Number of coreference chains that start with a third-person pronoun over the number of coreference chains. \\
\midrule

 &  Speech rate & -- & praat & Approximated number of syllables over total time~\cite{Feinberg_2022_praatscripts}. \\
 &  Articulation rate &  -- & praat & Approximated number of syllables over phonation time~~\cite{Feinberg_2022_praatscripts}. \\
  &  \raggedright  Average syllable duration &  -- & praat & Average syllable duration~~\cite{Feinberg_2022_praatscripts}. \\ 
  &  Mean pause duration &  -- & praat & Mean duration of silence segments, excluding silences before and after speech, motivated by~\cite{weiner2016alzheimer}.\\
 \centering \textbf{Rhythm}  &  Mean speech duration &  -- & praat & Mean duration of speech segments~\cite{weiner2016alzheimer}.\\
 &  Silence rate &  --  & praat & Total silence time over total time, motivated by~\cite{weiner2016alzheimer}.\\
  &  Silence-to-speech ratio &  -- & praat & Number of silent segments over the number of speech segments, motivated by~\cite{weiner2016alzheimer}.\\
  &  Mean silence count &  -- & praat & Number of silence segments over total time, motivated by~\cite{weiner2016alzheimer}].\\

\midrule
 & F0 & mean, std & praat & Fundamental frequency of vibration of the vocal folds. \\
 & HNR & -- & praat & Compares the energy in the harmonics to the energy in the non-harmonic (noisy) components of the speech signal~\cite{ritasingh2019profiling}. \\ 
 & local Jitter & -- & praat & Jitter refers to cycle-to-cycle perturbations of F0 in frequency. Speech with high jitter is perceived as roughness~\cite{ritasingh2019profiling}. Local jitter is the average absolute difference between consecutive periods, divided by the average period~\cite{praat_manual}, measured in \%.\\
 & local absolute Jitter & -- & praat & Average absolute difference between consecutive periods, measured in seconds~\cite{praat_manual}.\\
  \centering \textbf{Voice quality} & RAP Jitter & -- & praat & Relative average perturbation - the average absolute difference between a period and the average of it and its two neighbours, divided by the average period~\cite{praat_manual}.\\
 & ppq5 Jitter & -- & praat & Five-point Period Perturbation Quotient -- same as RAP jitter but based but computed with it and its four closest neighbours~\cite{praat_manual}. \\
 
 & local Shimmer & -- & praat & Shimmer refers to cycle-to-cycle variation of F0 in amplitude. Speech with high shimmer is perceived as buzzing~\cite{ritasingh2019profiling}. Local shimmer is the average absolute difference between the amplitudes of consecutive periods, divided by the average amplitude~\cite{praat_manual}, measured in \%. \\
 & local db Shimmer & -- & praat & Average absolute base-10 logarithm of the difference between the amplitudes of consecutive periods, multiplied by 20~\cite{praat_manual}. \\
 & apq3 Shimmer & -- & praat & Three-point Amplitude Perturbation Quotient -- average absolute difference between the amplitude of a period and the average of the amplitudes of its neighbours, divided by the average amplitude~\cite{praat_manual}. \\
 & aqpq5 Shimmer & -- & praat & Five-point Amplitude Perturbation Quotient -- same as apq3, but computed with it and its four closest neighbours~\cite{praat_manual}. \\
 & apq11 Shimmer & -- & praat & 11-point Amplitude Perturbation Quotient -- same as apq3, but computed with it and its ten closest neighbours~\cite{praat_manual}. \\
 
\midrule

 & F1 & mean, median & praat & Formants occur around frequencies that correspond to the resonances of the vocal tract. First formant frequency -- relates to the shape of the area behind the tongue (on the throat). If the ressonator has a small area, then the formant frequency should be higher. \\
\centering \textbf{Vocal tract} & F2 & mean, median & praat & Second formant -- relates to the shape of the area from the hump of the tongue to the tip of the lips \\
 & F3 & mean, median & praat & Third formant. \\
 & F4 & mean, median & praat & Fourth formant. \\

\bottomrule

\end{tabular}}
\end{table*}

\noindent
Singh~\cite{ritasingh2019profiling} distinguishes three processes for computational profiling of humans from their voice: knowledge-driven, data driven, or a combination of both. This work explores the latter. 
The mechanisms through which the different diseases impact speech, summarized in Fig.~\ref{fig:multi_diseases_impact_speech}, motivated the definition of a knowledge-driven feature set containing 41 interpretable features. Later, feature selection was conducted using a data driven approach.
This knowledge driven feature set contains both acoustic (28) and linguistic (13) features which are thoroughly described in Table \ref{tab:features_description}. We group the features into four categories: content-, rhythm-, voice quality- and vocal tract shape-related features. 
Content-related features were derived from automatically generated transcriptions. While the analysis of the picture description task encompassed all features, the analysis of sustained vowels was solely based on voice-quality and vocal-tract related features. 

Different methods were used to extract the features, as listed in Table \ref{tab:features_description}. The content-related features were extracted using the \textit{BlaBla} toolkit \cite{shivkumar2020blabla_feats}, dedicated scripts, or pretrained models. 
The coherence features were based on the cosine similarity between sentence embeddings of adjacent text segments, computed with the pretrained sentence-transformer model \textit{all-mpnet-base-v2}\footnote{available at \url{https://www.sbert.net/docs/pretrained_models.html}}. At the time this work was conducted, this model, trained with over 1 billion training pairs, provided the state of the art on the \textit{Sentence Embeddings Benchmark} \cite{reimers2019sBERT}. 
The embeddings were extracted for chunks of 14 tokens. The choice of 14 tokens was rooted on two reasons: (i) in CLAC, in the task where subjects describe the cookie theft picture, the average number of words per sentence in the provided transcriptions was 15, and in the task where subjects are describing the picnic picture, the average number of words per sentence is 13 words; and (ii) according to the~\cite{americanPressInst}, readers understand over 90\% of the information when sentences have 14 words.
After computing the cosine similarity of adjacent sentences, the mean and the variance are computed for the entire picture description. This measure of coherence was based on the \textit{incoherence model}, described in \cite{bedi2015psychosis} and \cite{iter2018schizophrenia} for the assessment of speech of subjects suffering from psychosis and schizophrenia. The use of the variance was inspired in the concept of ongoing semantic variability, proposed by \cite{sanz2022coherenceAD} as a text-level semantic marker of Alzheimer's Disease.

\textit{Ambiguous coreference chains} are sequences of words or phrases in a text that refer to the same entity or concept, which start with an ambiguous pronoun. Ambiguous pronouns refer to entities not explicitly mentioned or mentioned only cataphorically, i.e., after the pronoun. The usage of ambiguous pronouns, or referential incoherence, is a common pattern in incoherent speech. 
The usage of ambiguous pronouns was captured following the approach of \cite{iter2018schizophrenia}: (1) a pretrained coreference resolver extracts the reference chains (i.e., the lists of terms that should refer to the same entity), and (2) if the first term in the reference chain is a third-person pronoun (he, she, they, etc.), then it is considered an ambiguous pronoun. The pretrained coreference resolver was the wl-coref\footnote{available at \url{https://github.com/vdobrovolskii/wl-coref}.} \cite{dobrovolskii-2021-wl-coref}, that detained the state of the art on the CoNLL-2012 Shared Task\footnote{score board available at \url{http://nlpprogress.com/english/coreference_resolution.html}.}~\cite{pradhan-etal-2012-conll} at the time this work was conducted. The entire transcription of the picture description was used to compute this feature.

The remaining three feature categories -- rhythm, voice quality, and vocal tract shape -- were derived directly from the audio samples, using Praat~\cite{boersma2001praat}, through the Python package praat-parselmouth~\cite{parselmouth}.
Praat was chosen for its frequent use in clinical practice.

\vspace{10pt}
\noindent \textbf{Step 3. Outlier Removal}

\noindent
The identification of outliers in the reference population, i.e., samples that differ substantially from the other observations, was based on the Mahalanobis distance~\cite{mahalanobis2018generalized} to the mean of the population, an approach well-suited for multivariate data. The cutoff threshold was set to three times the standard deviation from the mean of the Mahalanobis distance. This analysis was conducted separately for the picture description and the sustained vowel tasks, using rhythm, voice-quality, and vocal-tract related features. Content features were not considered to ensure that results are independent of the ASR model.
 
Approximately 1\% of the audio samples in CLAC were considered outliers. By excluding these samples, we expect to exclude bad quality audios and possibly samples from speakers affected by subclinical diseases.
Table \ref{tab:ref_population} presents the number of audio samples and speakers in the reference population for each speech task after outlier removal.
 
\begin{table}[t]
\centering
\caption{Number of audio files and speakers, and average file duration in the reference population, per speech task, and by gender and age range, after vowel segmentation and outlier removal. \label{tab:ref_population}}
\vspace{-8pt}
\resizebox{0.95\linewidth}{!}{
\begin{tabular}{cccc|cc|cc}
\toprule
 & & \multicolumn{2}{c|}{CLAC$_{picture}$}  & \multicolumn{2}{c|}{CLAC$_{vowel}$} & \multicolumn{2}{c}{All}  \\
\cmidrule{3-8}
& & Files & Speakers & Files & Speakers & Files & Speakers \\
\midrule
\multicolumn{2}{l}{Count} & \\
\midrule
M & $<$50    & 1115 & 772 & 1040 & 598 & 2155 & 782 \\
M & $\geq$50 & 142  & 104 &  133 & 77  & 275  & 106 \\
\midrule
F & $<$50    & 1081 & 749 & 1044 & 641 & 2125 & 756 \\
F & $\geq$50 & 188  & 139 & 179  & 113 & 367  & 140 \\
\midrule
\multicolumn{2}{c}{All} 
            &  2526 & 1764 & 2396 & 1429 & 4922 & 1784 \\
\midrule
\multicolumn{4}{l}{Average duration $\pm$ standard deviation (s)} \\
\midrule
\multicolumn{2}{c}{All} 
            &  38 $\pm$ 22 & -- & 3 $\pm$ 1 & -- & -- & -- \\

\bottomrule
\end{tabular}
}
\vspace{-14pt}
\end{table}

\vspace{10pt}
\noindent \textbf{Step 4. Reference population partition}
\label{sec:multidisease-method-partition-ref}

\noindent
Most features in the feature set are strongly influenced by various factors, such as recording conditions~\cite{maryn2017mobile, jannetts2019assessing, dineley2023towards} and speaker dependent attributes, including gender, age, body mass index, accent, education, smoking habits, etc. 
The features may also exhibit substantial differences depending on the speech task (e.g. in spontaneous \textit{versus} read speech).
These factors could have a more substantial impact on the speech signal than speech-affecting diseases, potentially biasing results. 
It becomes important to determine whether different RIs should be estimated for different ranges of each of these factors.

Our previous work~\cite{botelho2023RIs} analysed when to partition the reference population and derive different RIs, based on gender, age range (above and below 50 years old), and speech task. This analysis involved assessing the statistical significance of differences between subgroups, by employing Mann-Whitney U tests~\cite{mcknight2010mann}.
The study recommended deriving distinct RIs for different genders and speech tasks. While different RIs for various age ranges are ideal, the data was deemed insufficient. Furthermore, to compare data from different source datasets, it is recommended that datasets be normalized separately, e.g. using zero-mean and unit-variance normalization.
Based on these findings, in this work, we partition the reference population by gender and speech task. 
These are simplifying assumptions, that we believe to be reasonable in this proof-of-concept exploring the feasibility of defining RIs for speech. Future work should not only study a larger reference population, but also consider other methods for partitioning the RIs, such as the Lahti criteria \cite{lahti2002partition}, or the Ichiahara method \cite{ichihara2010appraisal, ichihara2008sources}.

\vspace{10pt}
\noindent \textbf{Step 5. Reference intervals estimation}
\label{sec:c6-method-ref-int}

\noindent
In this work, an RI, i.e., the interval between the $2.5^{th}$ and $97.5^{th}$ percentiles~\cite{ichihara2010appraisal}, is derived for each feature, using the non-parametric approach.
Following the guidelines in~\cite{ozarda2016reference}, 90\% confidence intervals (CIs) were derived for both the lower and upper limits of the RI via boostrapping~\cite{Confidence_Intervals}, to provide a confidence measure on the estimated RI. Data was resampled 1000 times to estimate the confidence interval. 

We acknowledge that for certain features, it is more appropriate to provide a single boundary, either an upper or lower limit. For instance, elevated values of jitter and shimmer are considered pathological, possibly indicating affected laryngeal control, whereas there is no lower limit below which these values are deemed unhealthy. We believe that determining whether each feature should have a reference interval or a single limit should be guided by domain-specific knowledge, and we encourage further research on this topic. Our data-driven approach does not allow us to draw conclusions on this matter; therefore, reference intervals with two limits are derived for all features.

\vspace{10pt}
\noindent \textbf{Step 6. Feature correlation analysis}
\label{sec:c6-method-corr-analysis}

\noindent
Some of the features in the proposed feature set are very correlated with each other. It is expected, for instance, that the mean and median of the formants are very correlated with each other, and also measures of shimmer, and measures of jitter should be correlated amongst each other. Additionally, a high dimensionality feature space hinders the interpretability of the results by the medical community. 

Therefore, a feature correlation analysis is carried out to exclude redundant features and reduce the dimensionality.
Features were grouped into clusters of similar information using hierarchical clustering based on the Pearson correlation between all feature pairs. Different correlation thresholds, $CT\in\{0.5, 0.6, 0.7, 0.8, 0.9, 1\}$, were explored to fix the final clusters.
For $CT=1$, each cluster corresponded to a single feature. For other correlation thresholds, a \textit{"prototype feature"} was selected to represent each cluster, resulting in the final reduced-dimensionality feature set.

To promote robustness to dataset shifts, prototype features were chosen based on the similarity of the standard deviation of their distributions across different datasets. Ideally, other corpora of control individuals would be available for this analysis. Since that is not the case, the standard deviation of each feature in the reference population was compared to that of control subjects in the disease detection population. The feature with the most similar standard deviation in both groups within each cluster was designated the prototype feature. Means were not considered, as they can be adjusted by adding a bias term. Future work should explore more sophisticated methods for selecting prototype features.

This dimensionality reduction approach, which involves Pearson correlation analysis followed by hierarchical clustering on the reference population, was motivated by two primary objectives: first, to establish a feature set suitable for characterizing reference speech independently of disease-specific deviations; second, to mitigate the risks of unstable results, overfitting, and poor generalization associated with supervised feature selection on small datasets~\cite{dernoncourt2014analysis, soares2016feature, vabalas2019machine}.

Detailed results of the correlation analysis are provided in the next section.

%%%%%%%%%%%%%%%%%%%%%%%%%%%%%%%%
\subsection{\textbf{Reference Speech Results}}
\label{sec:results-task1}

\begin{figure*}[t]
  \centering
  \includegraphics[width=\linewidth]{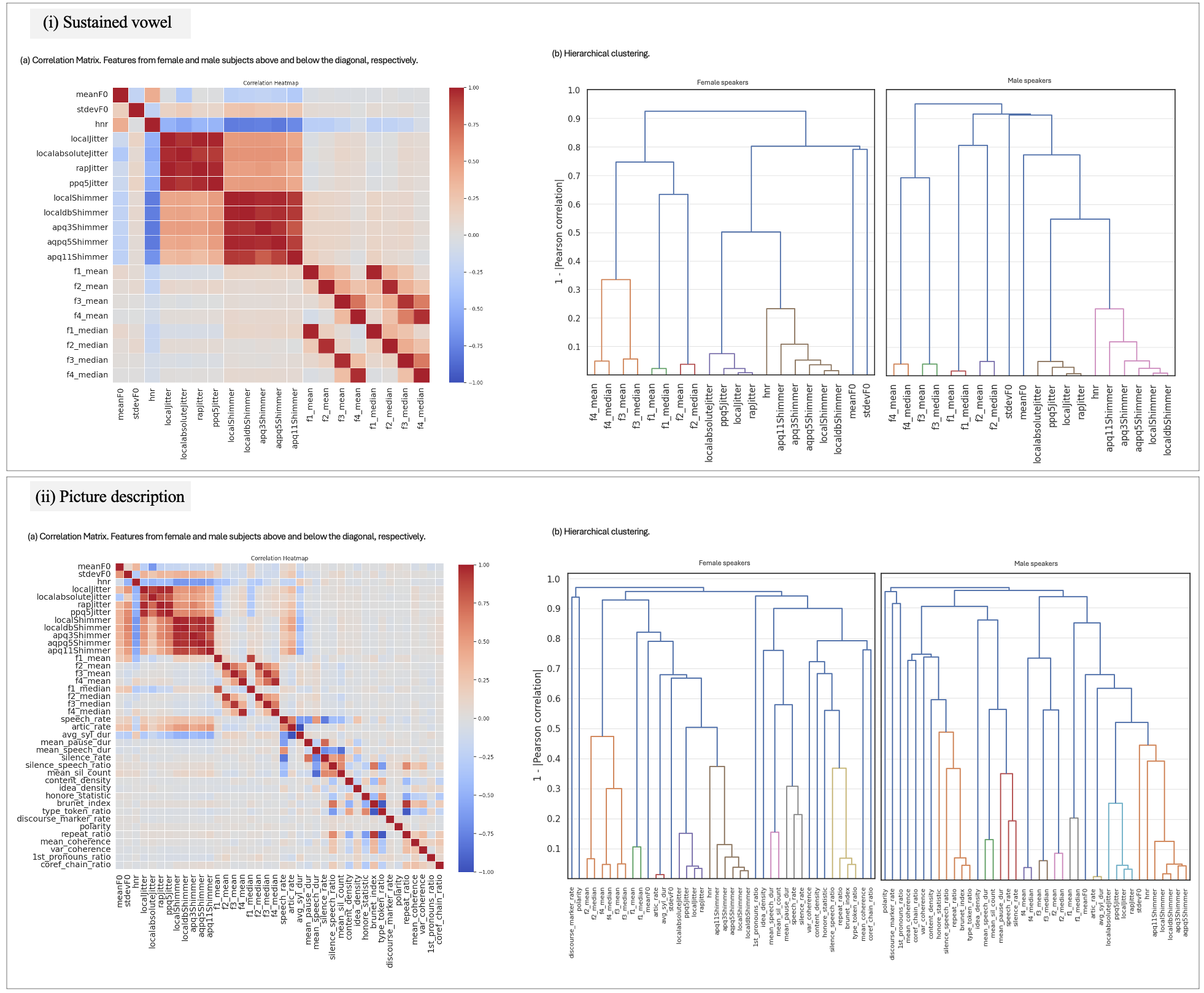}
  \vspace{-20pt}
  \caption{Correlation analysis of the features extracted from vowel recordings (top) and picture description transcriptions (bottom). (a) shows the Pearson correlation between the features. The values above the diagonal refer to features extracted for female subjects, while the values below the diagonal refer to male subjects. (b) shows the dendrogram results from the hierarchical clustering of features, based on their Pearson correlation correlation. The y axis corresponds to $1-CorrelationTreshold$, to capture the distance between features of the same cluster. 
  }
  \label{fig:feature_correlation_vowel_pic}
  \vspace{-8pt}
\end{figure*}

\begin{figure*}[t]
  \centering
  \includegraphics[width=\linewidth]{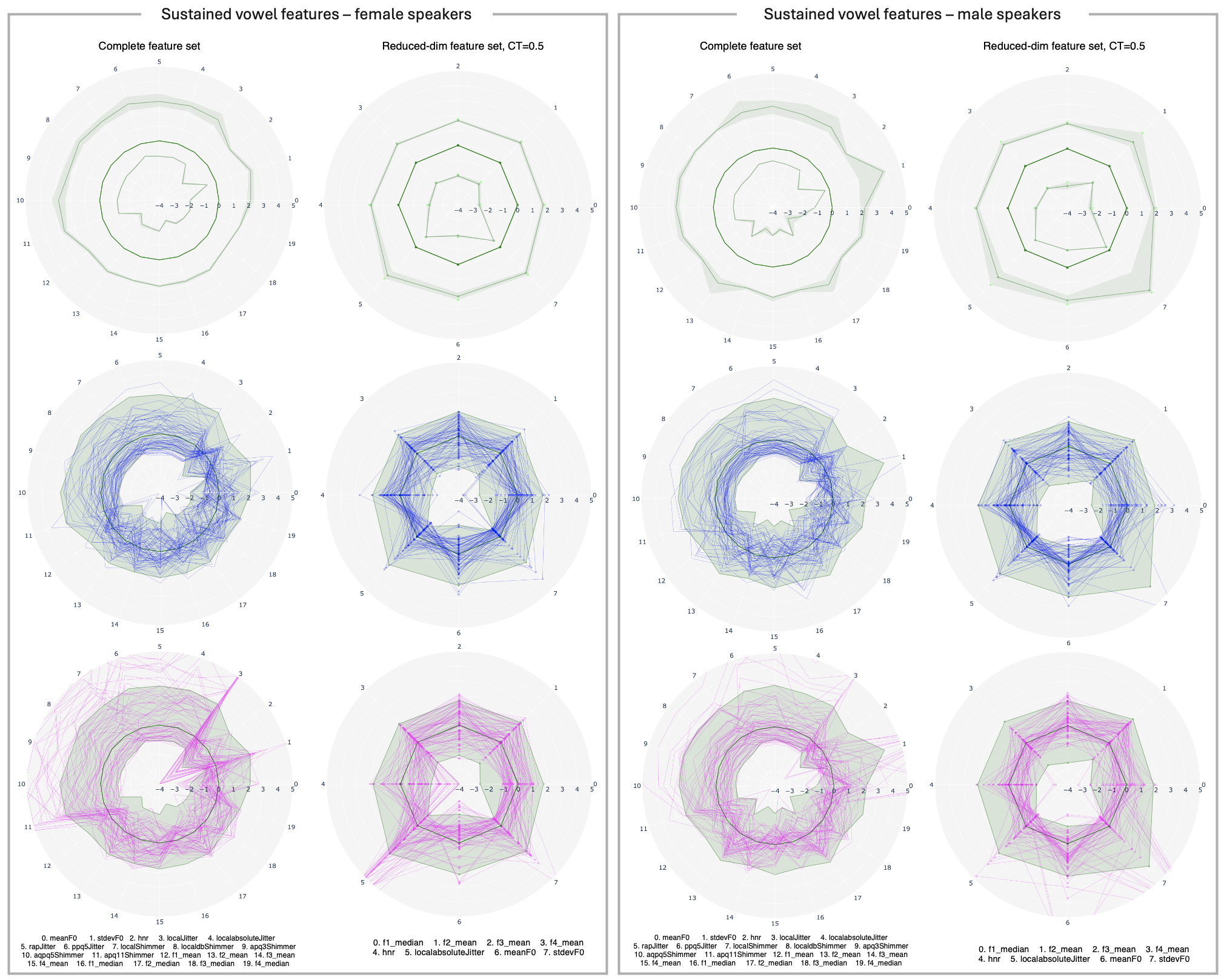}
  \vspace{-16pt}
  \caption{Radar plots to characterize reference speech, using the task \textit{sustained vowel /a/}. The dark green line corresponds to the mean value of each feature, while the light green lines correspond to the RI, computed using the reference population. Blue lines correspond to control speakers, whereas pink lines correspond to patients (PD).
  }
  \label{fig:radar_plots_vowels}
  \vspace{-10pt}
\end{figure*}

\begin{figure*}[t]
  \centering
  \includegraphics[width=\linewidth]{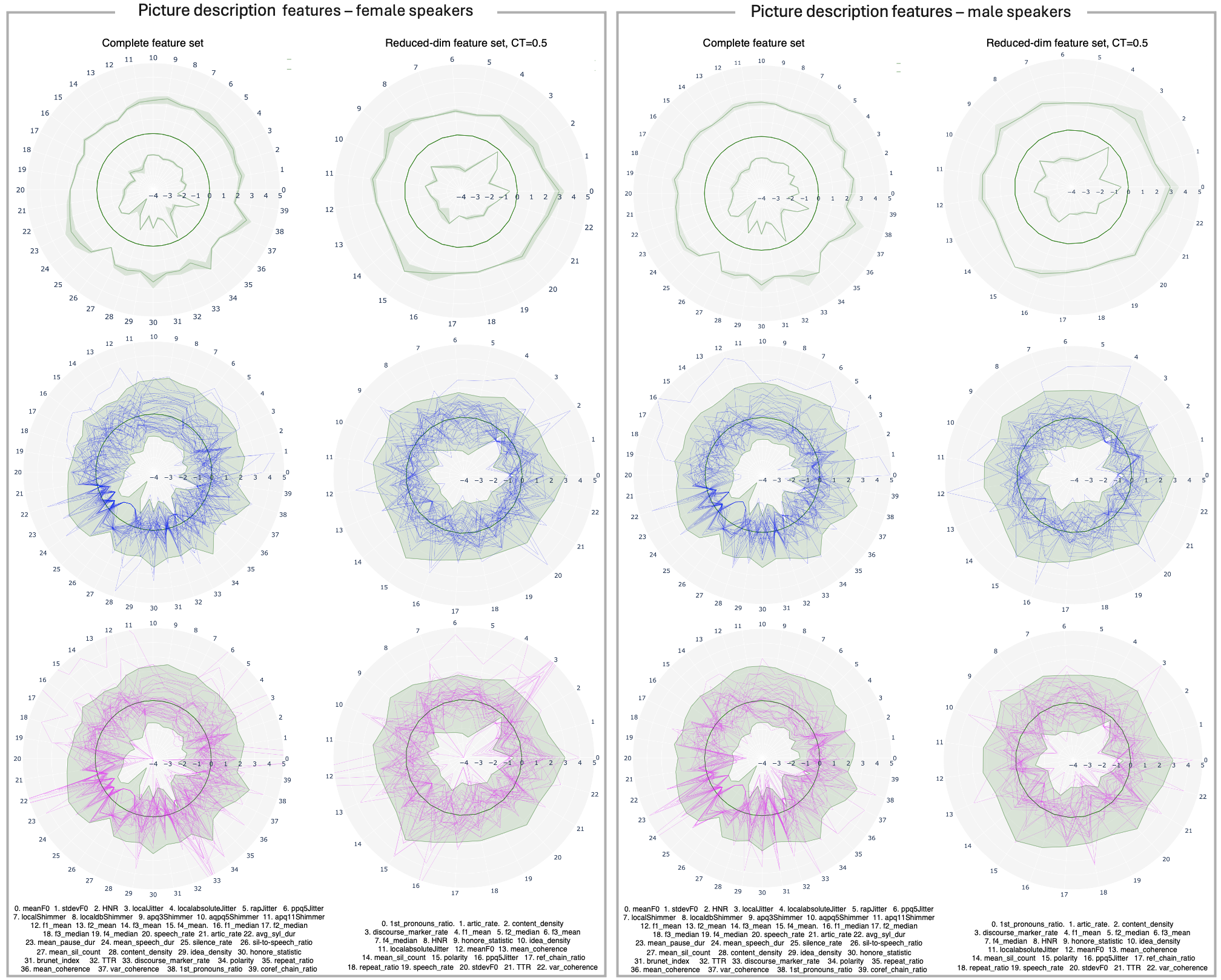}
  \vspace{-16pt}
  \caption{Radar plots to characterize reference speech, using the task \textit{picture description}. 
  The dark green line corresponds to the mean value of each feature, while the light green lines correspond to the reference interval, computed using the reference population. Blue lines correspond to control speakers, whereas pink lines correspond to patients (AD).
  }
  \label{fig:radar_plots_pic}
  \vspace{-12pt}
\end{figure*}

Reference intervals were derived for each feature in the full-dimensionality feature set\footnote{
Although the feature \textit{ratio of ambiguous coreference chains} may be interesting for the detection of several diseases, including Alzheimer's disease and schizophrenia, it was excluded from further analysis because the confidence intervals on the upper bound for both genders were larger than the RI itself, indicating a poor confidence on the derived RI. Further discussion is provided in Appendix~\ref{sec:appendix_ambiguous_coref_chain}.},
with the corresponding confidence intervals on the upper and lower bound. 
The analysis was conducted separately for the sustained vowel /a/ and the picture description task.

\vspace{10pt}
\noindent \textbf{Feature correlation analysis}

\noindent
As described in section~\ref{sec:c6-method-corr-analysis} (step 6), a reduced-dimensionality feature set was derived to exclude features highly correlated with each other. Fig.~\ref{fig:feature_correlation_vowel_pic} (a) shows a heatmap with the Pearson correlation values between all feature pairs. Strong colored cells correspond to pairs of strongly correlated features -- red for positively correlated features, and blue for inversely correlated features. The values above the diagonal correspond to features derived for female subjects, while the values below the diagonal correspond to values derived for male subjects. As expected, one can observe that mean and median values of formant frequencies are strongly correlated, as well as the different measures of jitter and different measures of shimmer.
HNR also appears inversely correlated with shimmer measures. Other patterns appear, for example speech rate is inversely correlated with silence rate, average syllable duration is inversely correlated with the articulation rate, and the repetition ratio is positively correlated with the Brunet's index and inversely correlated with the type-to-token ratio. Correlations between features are similar for both female and male subjects.

The clusters of highly correlated features achieved via hierarchical clustering are presented in Fig.~\ref{fig:feature_correlation_vowel_pic} (b).
The \textit{prototype-features} of each cluster, i.e., the features that were selected to represent the information encoded in each cluster, are listed in Table~\ref{tab:champion_feats}. The higher the correlation threshold ($CT$) that defines the clusters, the larger the number of clusters, and thus the higher the dimensionality of the final set.

\begin{table}[t]
\centering
\caption{{Prototype-features}, per correlation threshold, $CT$. \label{tab:champion_feats}}
\vspace{-8pt}
\setlength{\tabcolsep}{4pt}
\resizebox{\linewidth}{!}{
\begin{tabular}{p{1cm} p{10cm}} 
\toprule
\textbf{CT} & \textbf{Champion-features} \\
\midrule
\multicolumn{2}{l}{Sustained vowel} \\
\midrule

\textbf{0.5 / 0.6 / 0.7} & F1\_median, F2\_mean, F3\_mean, F4\_mean, HNR, localabsoluteJitter, meanF0, stdevF0 \\
\textbf{0.8} & apq11Shimmer, aqpq5Shimmer, F1\_median, F2\_mean, F3\_mean, F4\_mean, HNR, localabsoluteJitter, meanF0, stdevF0 \\
\textbf{0.9} & apq11Shimmer, aqpq5Shimmer, F1\_median, F2\_mean, F3\_mean, F4\_mean, HNR, localabsoluteJitter, localdbShimmer, meanF0, stdevF0 \\

\midrule
\midrule
\textbf{0.5 / 0.6} & First person pronouns, articulation rate, content density, discourse marker rate,
F1\_mean, F2\_median, F3\_mean, F4\_median, HNR, Honoré statistic, idea density, localabsoluteJitter, meanF0,
mean coherence, mean silence count, polarity, ppq5Jitter, coreference chain ratio, repetition ratio, speech rate, stdevF0, coherence variability\\
\textbf{0.7} & First person pronouns, articulation rate,
content density, discourse marker rate,
F1\_mean, F2\_median, F3\_mean, F4\_mean, F4\_median, HNR,
Honoré statistic, idea density, localabsoluteJitter,
localdbShimmer, meanF0, mean coherence,
mean pause duration, mean silence count, polarity, ppq5Jitter,
coreference chain ratio, repetition ratio, silence-to-speech ratio, speech rate,
stdevF0, TTR, coherence variability \\

\textbf{0.8} & First person pronouns, articulation rate,
content density, discourse marker rate,
F1\_mean, F1\_median, F2\_median, F3\_mean, F4\_mean, F4\_median, HNR,
Honoré statistic, idea density, localabsoluteJitter,
localdbShimmer, meanF0, mean coherence,
mean pause duration, mean silence count, polarity, ppq5Jitter,
coreference chain ratio, repetition ratio, silence rate, silence-to-speech ratio, speech rate,
stdevF0, TTR, coherence variability \\

\textbf{0.9} & First person pronouns, apq11Shimmer,
 articulation rate, 
 content density, discourse marker rate, F1\_mean, F1\_median,
 F2\_median, F3\_mean, F4\_mean, F4\_median, HNR,
 Honoré statistic, idea density, localabsoluteJitter,
 localdbShimmer, meanF0, mean coherence,
 mean pause duration, mean silence count, mean speech duration, polarity,
 ppq5Jitter, coreference chain ratio, repetition ratio,
 silence rate, silence-to-speech ratio, speech rate,
 stdevF0, TTR, coherence variability \\

\bottomrule
\end{tabular}
}
\vspace{-8pt}
\end{table}

\vspace{10pt}
\noindent \textbf{Comparison of reference and patient speech}

\noindent
Figs.~\ref{fig:radar_plots_vowels} and~\ref{fig:radar_plots_pic} show the RIs represented in a radar chart, normalized to zero mean and unit variance, for the sustained vowel task and picture description task, respectively.

In the top plots, light green lines indicate the lower and upper bound of the reference interval, with the shaded region representing the confidence interval on the reference interval limits. The dark green line represents the mean values, which are always zero due to normalization. The first and third column of each figure display the reference intervals for the entire feature set, while the second and forth columns refer to the feature subset obtained after the Pearson correlation-based feature selection, with $CT=0.5$. This reduced, less correlated, feature set aims to highlight which groups of features are more impacted by each disease.
The speech of any subject, while performing one of the speech tasks analysed, can be projected into the radar plot, and compared to the reference population. Ideally, if a subject is healthy, their speech should be represented within the area delimited by the reference intervals.

On the second and third row plots, we overlay individual data from the disease detection population onto the reference interval radar charts.
Following the discussion in section~\ref{sec:multidisease-method-partition-ref}--step 4, the population for disease detection was also normalized to zero-mean and unit variance, using only the control subjects to compute the statistics for normalization.
Each subject is represented by a different line, control subjects in blue and patients in magenta, shown in separate plots. The entire area within the reference interval is shaded to enhance visibility. 

By visual inspection, the confidence intervals on the limits of the RIs derived for the sustained vowel task (Fig.~\ref{fig:radar_plots_vowels}) appear relatively narrow, with the exception of the \textit{standard deviation of F0} and the \textit{mean of the second formant} for male subjects. For the picture description task (Fig.~\ref{fig:radar_plots_pic}), the feature with a wider confidence interval is \textit{silence-to-speech ratio}. One can interpret these wide confidence intervals as an indication of lack of confidence on the exact margins of the RI. Future research should aim at improving the confidence of these intervals with a larger reference population, collected under controlled recording conditions. 

When visually inspecting the plots representing the PD patients enunciating a sustained vowel /a/ (Fig.~\ref{fig:radar_plots_vowels}), it is clear that they deviate from the reference values more frequently than controls in the axis that correspond to HNR, jitter, shimmer and F0 related features. This is notorious for both genders, however it appears that HNR, jitter and shimmer features are more relevant for female subjects, while F0 features are more relevant for male subjects.

When analyzing the Cookie Theft image descriptions (Fig.~\ref{fig:radar_plots_pic}), differences between AD patients and controls also exist. 
For instance, several female AD patients exhibit a discourse marker rate (feature 33) substantially above the reference interval. Specifically, 26\% of female AD patients surpass the RI for discourse marker rate, compared to only 7\% of female controls. Additionally, speech rate (feature 20) for both male and female AD patients is more frequently below the RI than that of control subjects.

However, these differences appear more notorious in the sustained vowel task than in the picture description. This may indicate that the task of enunciating a sustained vowel may be more suitable for this RI analysis, as it entails less sources of variability. It is also possible that the noisy recording conditions in ADReSS play a strong role.

This radar chart visualization is particularly well-suited for the analysis of speech as a biomarker for health in two scenarios.
When studying a disease population, the radar chart visualization enables the identification of features that appear to be strong markers of a disease, and simultaneously still robust to dataset shifts. Taking the example of PD female patients \textit{vs} Control females (Fig. \ref{fig:radar_plots_vowels} -- left): there are 6 features (all jitter-, shimmer- and HNR-related) for which more than 95\% of the controls stay inside the RI and over 20\% of the patients fall outside the RI. 
Alternatively, this visualization provides a simple way to compare the speech features of one individual to the speech features of a reference population, and quickly identify if there are any deviations on groups of features that are expected to be affected by a certain disease. 

Radar plots have been previously employed to visualize speech features in the context of speech as a biomarker. Jiao et al.~\cite{jiao_berisha2017interpretable} used radar plots to illustrate phonological disturbances in dysarthric speakers, while Behrendt et al.\cite{tese2023Behrendt} introduced DemVis, a prototype system for extracting and visualizing speech features, which also performs AD detection.

A quantitative comparison of disease-affected speech with the RIs derived for the reference population was conducted, namely to compare the distance of the features in disease-affected speech to the reference intervals. This distance is, on average, higher for patients than for controls, and the difference between the two groups is considered statistically significant for female and male subjects, both in PC-GITA and ADReSS. Further details are available in Appendix~\ref{sec:appendix_ri_quantitative_analysis}.

%%%%%%%%%%%%%%%%%%%%%%%%%%%%%%%%%%%%%%%%%%%%%%%%%%%%%%
\section{Classification of speech affecting diseases}
\label{sec:task2-disease-detection}

\begin{figure*}[t]
  \centering
\includegraphics[width=0.99\linewidth]{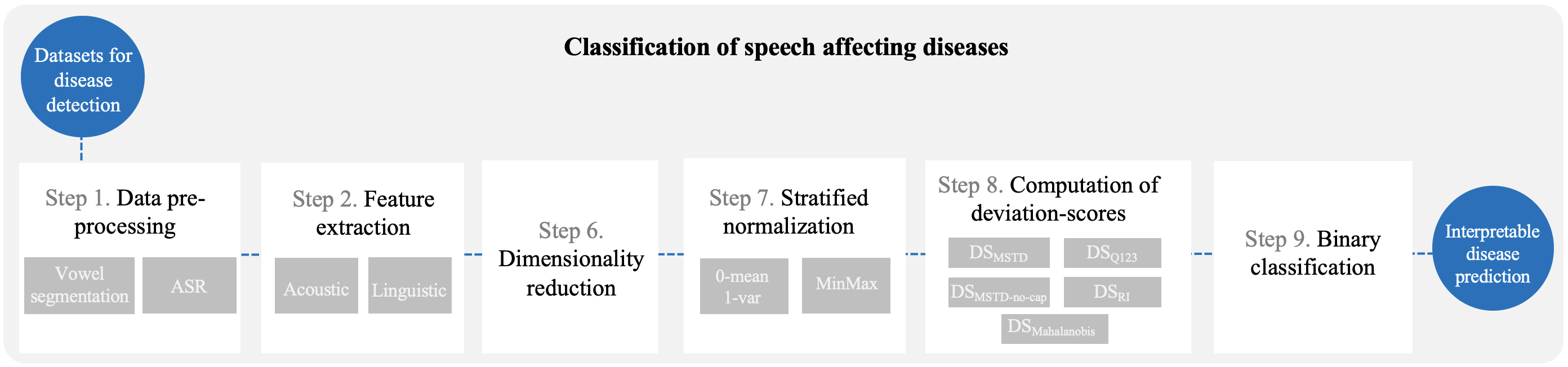}
\vspace{-8pt}
  \caption{Overview of the steps entailed in the detection of diseases.}
  \label{fig:pipeline_task_2}
  \vspace{-10pt}
\end{figure*}

This section focuses on the automatic detection of speech-affecting diseases leveraging the definition of reference speech described before.
\textit{Deviation-scores} are introduced to quantify how much an individual's speech feature deviates from reference values. These scores are then used as inputs for the detection task.
The detection of each speech affecting disease is formulated as a binary classification problem (patients versus controls) and addressed using Neural Additive Models (NAMs)~\cite{agarwal2021nams}. NAMs provide full transparency, enabling explanations that are compatible with clinical reasoning.

This analysis focuses on the detection of Alzheimer's disease and Parkinson's disease separately, but ultimately it is aimed at defining an approach suitable for multiple speech-affecting diseases. In fact, we propose a single framework that could be used for the detection of several diseases, differing slightly on the subsets of features to be used, depending on the speech task at hand, and on the model used for classification. As described earlier in section ~\ref{sec:task1-ref-speech-method} (step 2), for the picture description task, all feature groups should be used, while for sustained vowels, only \textit{voice quality} and \textit{vocal tract} related features are suitable. We expect that future work analysing reading tasks should include all feature groups, except content related features, and spontaneous speech tasks should leverage all feature groups.

\subsection{\textbf{Disease detection pipeline}}

The proposed method entails several steps, some of them -- pre-processing, feature extraction, and dimensionality reduction -- shared with the previous stage for reference speech definition, as depicted in Fig.~\ref{fig:pipeline_task_2}.
The vowel segmentation step (section~\ref{sec:c6-method-vowel-processing}--step 1) led to the exclusion of 12 PC-GITA samples from PC-GITA (further details in Appendix~\ref{sec:appendix_data_preprocess_feature_ext}). These samples were not used for classifier training or hyperparameter optimization. 
However, to ensure a fair comparison with previous literature, these excluded samples are arbitrarily assigned the prediction "control" when reporting performance on the test set. This reflects a 50\% a priori probability of a correct prediction due to the balanced nature of the datasets.

After data pre-processing, feature extraction, and dimensionality reduction, data underwent normalization (step 7), and deviation scores were computed (step 8). 
This process culminated in binary classification (Step 9), which outputs an interpretable prediction. Detailed descriptions of these steps are provided below.
While steps 3 to 5 pertain exclusively to the speech characterization pipeline and are not part of the disease detection pipeline, the numbering is retained across both pipelines for consistency, reflecting the shared nature of some steps.

\vspace{10pt}
\noindent \textbf{Step 7. Stratified normalization}
\label{sec:method_normalization}

\noindent
Prior to comparing reference intervals with disease datasets and calculating deviation scores, stratified normalization by gender and source dataset was performed. This approach preserves the intrinsic characteristics of each group, acknowledging that the relationship between controls and patients may differ between male and female subjects, and that the distribution of each feature may differ in each dataset. 
Only the control subjects in the training set within each stratification group were used to compute the statistics for normalization.
We compared two normalization strategies, zero mean and unit variance scaling, and \textit{MinMax} scaling between zero and one. Both strategies were implemented using the scikit-learn toolkit~\cite{scikit-learn}.
This approach assumes the gender of speakers in the test set is known, a reasonable assumption given the high performance of automatic speech-based gender detection methods~\cite{kwasny2021gender}.

\vspace{10pt}
\noindent \textbf{Step 8. Computation of deviation-scores}

\noindent
In the previous stage, reference speech was characterized using the distribution of acoustic and linguistic features within a reference population. The hypothesis explored here is that deviations of a new audio sample relative to the reference population can indicate the presence of a specific disease in the speaker. Five deviation scores ($DS$) were compared to assess the extent to which each feature value \(x_i\) in a new audio sample diverges from the corresponding feature distribution in the reference population. The five deviation scores are: $DS_{MSTD}$, inspired by ~\cite{zusag23_carefulwhisper}, $DS_{MSTD-no-cap}$, $DS_{Q123}$, $DS_{RI}$, and $DS_{Mahalanobis}$. Further details on these deviation scores, together with exhaustive experiments using them are reported in the Appendices.

\vspace{10pt}
\noindent \textbf{Step 9. Binary classification}
\label{sec:method_classification}

\noindent
Classification experiments were conducted with Neural Additive Models (NAMs)~\cite{agarwal2021nams}, a type of glass-box models inherently interpretable. NAMs are part of the model family called Generalized Additive Models (GAMs), which are described by

\vspace{-20pt}
\begin{equation}
\label{eq:gams}
    g(\mathbb{E}[y]) = \beta + f_{1}(x_{1}) + f_{2}(x_{2}) + \cdots + f_{K}(x_{K})\,, 
\end{equation}

where  x$=(x_{1}, x_{2}, ..., x_{K})$ is the input with $K$ features, $y$ is the target variable, $g$ is the link function, and $f_{i}$ is a univariate shape function with $\mathbb{E}[f_{i}]=0$. 

The idea of NAMs is to parameterize each $f_{i}$ in Equation~\ref{eq:gams} by a neural network (subnet). In short, NAMs  are a linear combination of neural networks, each attending to a single feature, that are trained jointly using backpropagation. It is this modularity that makes NAMs' predictions very easily interpretable. NAMs' predictions can be interpreted by visualizing each of the learned shape functions, e.g. plotting $f_{i}(x_{i})$ \textit{vs} $x_{i}$. The graphs learnt by NAMs  are not \textit{a posteriori} explanations, but rather an exact description of how the model comes to a prediction.

The NAM architecture is also compatible with a multitask-scenario, particularly suitable for the simultaneous detection of multiple speech affecting diseases, when adequate data is available. The multitask architecture is identical to that of single task NAM, except that each feature is associated with multiple subnets and the model jointly learns a task-specific weighted sum over their outputs that determines the shape function for each feature and task~\cite{agarwal2021nams}. We expect that this property will be key to enable the simultaneous detection of multiple diseases, provided that datasets with similar tasks collected under comparable conditions are available.

The binary classification experiments were based on ADReSS for Alzheimer's disease detection, and PC-GITA for Parkinson's disease detection. Given the limited size of these corpora, the experiments were conducted in a 10-fold cross validation (CV) setting. 
For ADReSS, because a held-out test set was defined for the challenge in which the corpus was introduced, the 10-fold cross validation was applied on the training set, for hyperparameter tuning. 
Afterwards, the predictions for the test set of the 10 models trained during CV are aggregated via majority voting. 
For PC-GITA, there was no held out test set, hence the 10-fold CV was conducted on the entire dataset. In each run, one of the 9 training folds was assigned as development fold, to perform hyperparameter tuning.
Folds were defined to ensure that all data from the same speaker is assigned to the same fold, to avoid leakage of speaker information across training, development and test folds. Folds ensure a balance between healthy controls and patients, in terms of number of speakers, gender and age. 

Data was normalized separately for both genders, as described in section~\ref{sec:method_normalization}--step 7, with the normalization statistics being computed based on the controls in the training folds.
Distance scores were also computed separately for both genders, given that the reference intervals were derived separately for both genders. The classifiers, however, are gender-independent.

Initially, an exhaustive set of binary classification experiments were conducted with Support Vector Machines (SVMs) and Logistic Regression, using the two transcription types, three normalization strategies (zero-mean and unit variance, \textit{MinMax}, and no normalization), six correlation threshold values, and five deviation-scores were compared. An extra scenario where the raw features are directly fed to the classifiers after the stratified normalization was also considered. The results obtained are thoroughly described in Appendix~\ref{sec:appendix_results_svm_lr}.
The configurations that yielded the best results
were used to restrict the range of experiments conducted with NAMs, given their higher computational burden. Hyperparameter tuning was performed with Bayesian optimization using Gaussian Processes, implemented on scikit-optimize~\cite{Head2021scikit-optmize}, with 100 calls to the optimizer. Further details on training hyperparameters and network architecture are reported in the Appendix~\ref{sec:appendix_nams_architecture}.

%%%%%%%%%%%%%%%%%%%%%%%%%%%%%%%%%%%%%%%%%%%%%%%%%%%%%%%%%%%%%%
\subsection{\textbf{Disease detection results}} 
\label{sec:results}

\begin{table*}[t]
  \caption{Classification results, using NAMs, in terms of accuracy (Acc), macro precision (P), macro recall (R), and macro F1, in [\%].}
  \vspace{-7pt}
  \label{tab:results_nam_complete}
  \centering
  \resizebox{0.85\linewidth}{!}{
  \begin{tabular}{cccc cccccccccccc}
    \toprule
& & & & \multicolumn{4}{c}{\textbf{Dev Folds}} & \multicolumn{4}{c}{\textbf{Test}} & \multicolumn{4}{c}{\textbf{Test -- Speaker MV}} \\
\textbf{CT} & \textbf{DT} & \textbf{norm} & \textbf{Acc} & \textbf{P} & \textbf{R} & \textbf{F1} & \textbf{Acc} & \textbf{P} & \textbf{R} & \textbf{F1} & \textbf{Acc} & \textbf{P} & \textbf{R} & \textbf{F1} \\
 \midrule
\multicolumn{2}{l}{\textbf{Parkinson's Disease}} \\
\midrule																
CT=1.0 & $DT_{RI}$ & MinMax & \textbf{75.0} & \textbf{75.8} & \textbf{74.9} & \textbf{74.8} & \textbf{68.7} & \textbf{69.9} & \textbf{68.7} & \textbf{68.2} & \textbf{73.0} & \textbf{74.7} & \textbf{73.0} & \textbf{72.5} \\
CT=0.9 & $DT_{Q123}$ & MinMax & 72.6 & 73.2 & 72.5 & 72.4 & 62.7 & 64.0 & 62.7 & 61.8 & 67.0 & 69.2 & 67.0 & 66.0 \\
CT=1 & $DT_{Q123}$ & MinMax & 72.2 & 73.0 & 72.2 & 72.0 & 66.3 & 67.3 & 66.3 & 65.9 & 68.0 & 69.1 & 68.0 & 67.5 \\

\midrule
\multicolumn{2}{l}{\textbf{Alzheimer's Disease}} \\
\midrule
CT=1.0 & Raw feats & MinMax & 83.3 & 83.4 & 83.3 & 83.3 & 72.9 & 73.3 & 72.9 & 72.8 & -- & -- & -- \\ 
CT=0.7 & $DT_{RI}$ & none & 73.1 & 75.7 & 73.1 & 72.5 & 70.8 & 70.8 & 70.8 & 70.8 & -- & -- & -- & -- \\ 
CT=0.5 & $DT_{RI}$ & MinMax & 79.6 & 80.3 & 79.6 & 79.5 & 70.8 & 72.2 & 70.8 & 70.4 & -- & -- & -- & -- \\ 
CT=0.5 & Raw feats & MinMax & \textbf{84.3} & \textbf{84.4} & \textbf{84.3} & \textbf{84.2} & \textbf{75.0} & \textbf{75.0} & \textbf{75.0} & \textbf{75.0} & -- & -- & -- & -- \\ 
CT=0.5 & $DT_{RI}$ & none & 75.9 & 77.8 & 75.9 & 75.5 & 72.9 & 73.0 & 72.9 & 72.9 & -- & -- & -- & -- \\ 
CT=1.0 & Raw feats & none & 81.5 & 81.9 & 81.5 & 81.4 & 72.9 & 73.0 & 72.9 & 72.9 & -- & -- & -- & -- \\ 
\bottomrule
\end{tabular}
}
\end{table*}

\begin{figure*}[t]
  \centering
\includegraphics[width=0.9\linewidth]{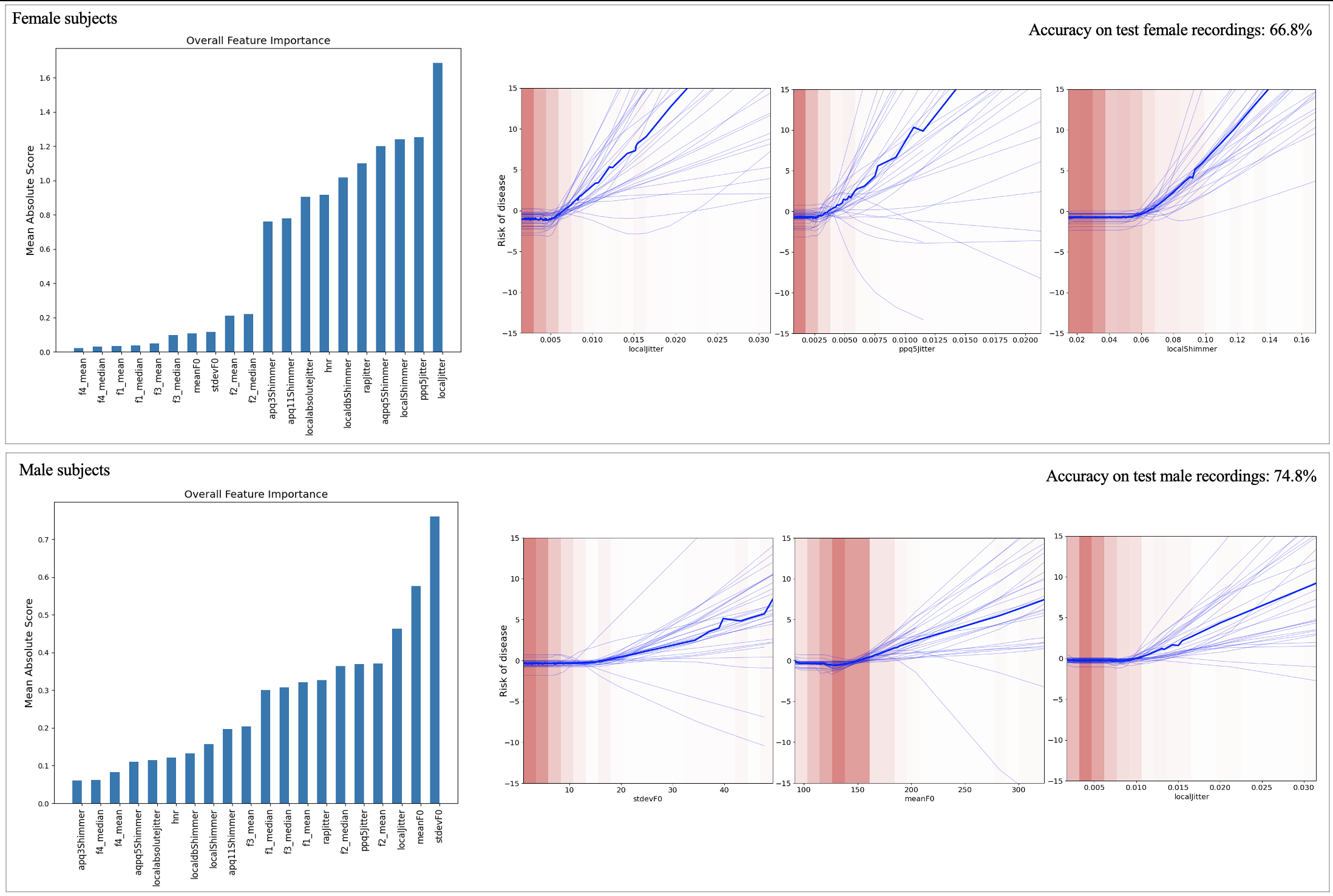}
  \caption{NAM trained for PD classification, for female (top) and male (bottom) subjects. The left plots represent the features that most contribute to the predictions; the plots on the right depict the shape functions learnt by the NAM, for the top three most important features.}
  \label{fig:nams_pcg}
  \vspace{-12pt}
\end{figure*}

\begin{figure*}[t]
  \centering
\includegraphics[width=0.9\linewidth]{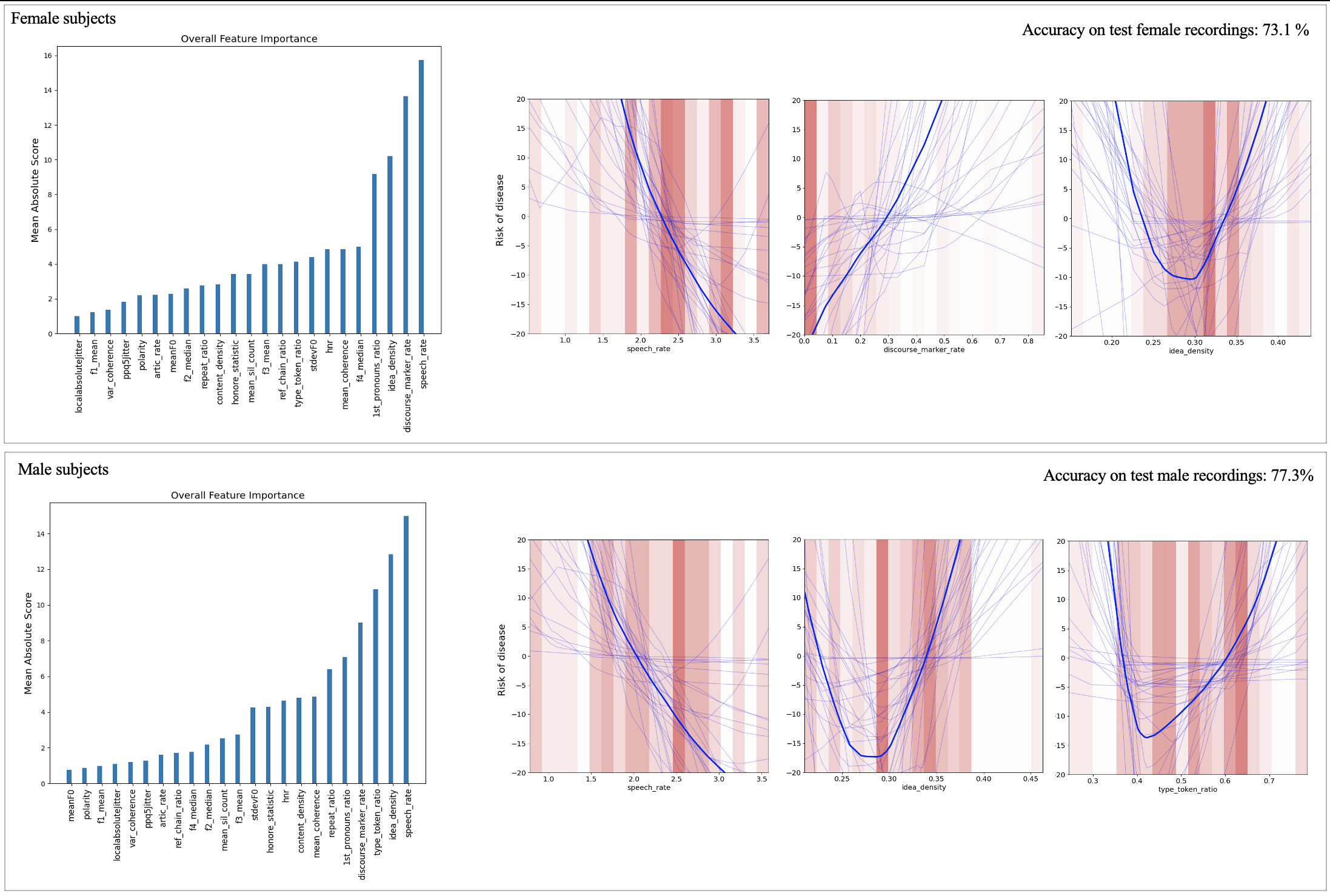}
  \caption{NAM trained for AD classification, for female (top) and male (bottom) subjects. The left plots represents the features that most contribute to the predictions; the plots on the right depict the shape functions learnt by the NAM, for the top three most important features.}
  \label{fig:nams_adress}
  \vspace{-6pt}
\end{figure*}

The results obtained with NAMs on the different configurations explored are reported in Table~\ref{tab:results_nam_complete}.
For PD classification, NAMs were able to achieve 75\% accuracy on the development folds, 69\% of the test folds, and a speaker-level accuracy after majority vote of 73\%. For AD detection, NAMs yielded 84\% on the development folds, and 75\% on the held-out test set. 
The NAM performance for PD detection on the test folds is lower than that achieved with LR/SVM classifiers. Conversely, for AD detection, the NAM results are better that those obtained with LR/SVM classifiers (based on the best development performance). The best classification results for PD were obtained with the entire feature set, while for AD were obtained with the reduced dimensionality feature set, with $CT=0.5$, similarly to the SVM and LR results (detailed in Appendix~\ref{sec:appendix_results_svm_lr}). Notably, the dimension of the entire feature set used to study sustained vowels (20 features) is very similar to the dimension of the reduced set used for studying the picture description task (23 features).

More importantly than surpassing the classification performance, the advantage of NAMs lies in their inherent interpretablity. The modularity and transparency of NAMs allow for precise visualization of what the models learn during training and how each prediction is computed.
Figs.~\ref{fig:nams_pcg} and \ref{fig:nams_adress} depict the features that contribute most to the predictions, and present the corresponding learned shape functions. These graphs are not \textit{a posteriori} explanations, but rather an exact description of how the model comes to a prediction.
Each semi-transparent blue line corresponds to one model of the ensemble, trained on a given run of the cross-validation, i.e. the lines correspond to 30 models (an ensemble of 3 models was trained for each of the 10 runs in the 10-fold cross-validation, as detailed in Appendix~\ref{sec:appendix_nams_architecture}). The solid blue line corresponds to the average of each model. Following~\cite{agarwal2021nams}, the average score of each feature (each shape function) was set to zero, by subtracting the mean score, averaged over the entire training set. This results that, on binary classification tasks, positive scores increase the probability of the positive class, compared to the baseline probability of observing that class, while negative scores decrease the probability. On the same plots, the normalized data density is also visible, in the form of pink bars. The darker the shade of pink, the more data there is in that region.

One can observe substantial variability in the shape functions learnt by each model, particularly when less data is available for training, indicated by lighter shades of pink. This is very evident in the ADReSS corpus, which is half the size of PC-GITA.
This variability gives a sense of confidence on the patterns learnt and emphasizes the need for research with larger corpora to enable more robust conclusions.
Nevertheless, in most cases, the models learned curves that align with the expected manifestations of each disease in the speech signal.

Upon analysing the NAM trained for PD detection which achieved the best accuracy on the test set (Fig.~\ref{fig:nams_pcg}), it becomes evident that the outcomes differ between genders, although 
\textit{local jitter} appears on the top three most important features for both genders.
This feature is, by far, the most important for female speakers. The following features in terms of importance for female speakers are also jitter and shimmer-related. This is consistent with expectations, as a high jitter and shimmer reflect cycle-to-cycle perturbations of F0, associated with impaired laryngeal control typical of PD patients. Previous studies have also found higher jitter values in PD patients when compared to control subjects (e.g.~\cite{jimenez1997acoustic, upadhya2017statistical}).

For male subjects, the most important features are the standard deviation and mean of F0. The corresponding shape functions indicate that higher values of mean F0 and of F0 standard deviation are associated with a higher risk of PD. 
Other works have also found increased F0 standard deviation in sustained vowels produced by PD patients compared to control subjects (e.g.~\cite{goberman2002phonatory, midi2008voice}). Goberman et al.~\cite{goberman2002phonatory} suggested that this increase may be due to laryngeal instability, potentially caused by weakness of the laryngeal musculature resulting from rigidity or tremor. The author mentions that tremor-related weakness has been found in other body systems, such as wrists~\cite{brown1997does}. 
Although not studied here, it is noteworthy that, in continuous speech, as opposed to sustained vowels, the F0 standard deviation is expected  to decrease in PD patients, typically described as mono-pitch~\cite{bowen2013effects, ma2020voicePD, harel2004variability}. 

The mechanism underlying the increased mean F0 in PD patients is suggested to be the increased rigidity of the laryngeal musculature (e.g. cricothyroid and thyroarytenoid muscles)~\cite{duffy1995speech, ma2020voicePD}. Biomechanical models of phonation demonstrate that increased vocal fold stiffness leads to higher fundamental frequency and jitter~\cite{ma2020voicePD}. Various works have identified differences in mean F0 in PD patients and controls, although not always significant for both genders, nor in the same direction. For example, Goberman et al.~\cite{goberman2002phonatory} found that mean F0 was higher in PD patients than controls, particularly in male speakers; Midi et al.~\cite{midi2008voice} found that mean F0 was higher in PD patients than healthy controls, but this difference was only significant in female subjects; and Yang et al.~\cite{yang2020physical} found the opposite, i.e., that mean F0 was lower in patients suffering from PD than controls.
It is important to acknowledge that F0 is more than just a marker for vocal fold behaviour, it carries information about different speaker states and traits~\cite{cummins2015analysis}, and even physiological aspects such as hormonal balance and aging~\cite{ritasingh2019profiling}. Thus, results should be interpreted with caution, and further research should be conducted with a larger dataset of age- and gender-matched controls and patients of PD and other diseases.

The patterns learnt by the NAM trained for PD detection are consistent with those represented on the radar plot of the sustained vowel for PD patients (Fig.~\ref{fig:radar_plots_vowels}). Both the NAM and radar plot flag jitter and F0 as important features for characterizing the speech of female and male PD patients.

Upon examination of the NAM trained for Alzheimer's disease detection which achieved the best performance, as depicted in Fig.~\ref{fig:nams_adress}, it is evident that each model in the ensemble/cross validation run learnt different patterns, yet some general trends can be inferred when averaging the predictions of those models. The first consideration is that the foremost contributing feature to prediction, for both male and female subjects, is \textit{speech rate}. It is clear that the slower the speech rate, the higher the risk of the person suffering from AD. This behaviour is expected, as a slower speech with more pauses is expected to be associated with a higher risk of AD. In fact, other works (e.g. ~\cite{hoffmann2010temporal_ad, hecker2022voice}) have also identified the importance of speech rate for AD detection from speech.

\textit{Idea density} is also among the top three most important features for detecting AD in both female and male speakers. The shape function learnt reflects a U-shape, indicating that low and high values of the feature idea density are associated with a higher risk of Alzheimer's disease. 
Low idea density has been associated with Alzheimer's disease at least since the well-known "Nun Study" by by Snowdon et al.~\cite{snowdon1996linguistic}, which found that low idea density in early life strongly predicted reduced cognitive ability or the presence of AD in later life. Boschi et al.~\cite{boschi2017connected} conducted a comprehensive review and reported that AD patients have significantly lower idea density compared to controls.

A similar U-shape pattern was learnt by the subnet attending to the Type-to-token ratio (TTR) feature, which is one of the top-three most important features for AD detection in male subjects.
TTR, a measure of lexical diversity, is also associated with cognitive impairments. For example, Bucks et al.~\cite{bucks2000analysis} found TTR to be significantly lower in AD patients compared to control subjects. Berisha et al.~\cite{berisha2015tracking} suggested that measures of lexical diversity, including the number of unique words, are strong predictors of pre-clinical AD onset.

Contrarily, the fact that the NAM learnt to associate high values of these features with a higher risk of AD is more surprising, and not frequently reported in the literature, to the best of our knowledge. However, this pattern observed by the NAM is present in the data.
We manually inspected data samples labelled with AD associated high idea density and/or TTR and identified several patterns: some examples, although not very frequent, included whisper hallucinations; some examples corresponded to correct transcriptions,  but were confusing or nonsensical despite high idea density or TTR; and occasionally, they were perfectly coherent descriptions of the Cookie Theft picture. 
These findings reinforce the idea that  features such idea density or TTR alone are not sufficient to make a prediction. 

Nevertheless, this illustrates the advantage of having a fully transparent model, despite its imperfections. For instance, let us consider a scenario where a new individual undergoes testing with this system and receives a prediction of Alzheimer's disease. A healthcare practitioner could examine the reasons provided by the system for this prediction. If the sole reason provided was a high idea density or high TTR, the healthcare practitioner would have the information needed to make an informed decision or recommend further testing.
Such reasoning would not be possible with a black-box model, or with a model that operates on uninterpretable features.

Finally, the \textit{Discourse marker rate} is also among the top-three most influential features for AD detection in female subjects. As the discourse marker rate increases, there is a higher risk associated with Alzheimer's Disease, consistent with the findings of Boschi et al.~\cite{boschi2017connected}.

%%%%%%%%%%%%%%%%%%%%%%%%%%%%%%%%%%%%%%%%%%%%%%%%%%%%%%%
\section{Limitations}
\label{sec:limitations}

This exploratory work has some limitations, and further research is needed to address these potential drawbacks.
One obvious limitation concerns feature extraction. Some of the features used, particularly vocal tract and voice-quality features, have been noted for their limited robustness across diverse recording conditions, including various devices--especially mobile platforms--background noise, and reverberation~\cite{maryn2017mobile, jannetts2019assessing, dineley2023towards}. 
Therefore, these features may not consistently yield reliable results across corpora recorded under different conditions. We advocate for the need to come up with guidelines to standardize how researchers record corpora and extract these features, which would enhance robustness and facilitate fair comparisons among different studies.

Another limitation pertains to corpora availability. The CLAC dataset, while valuable and larger than most speech corpora in clinical research, is crowdsourced, resulting in noise and lack of medical verification despite data filtering. Additionally, its size conditions the reliability of results, particularly for RI estimation, which ideally requires a minimum of 400 subjects per gender and age range. The small size of PC-GITA and ADReSS also adversely affects disease detection, specifically impacting the shape functions learned by NAMs' subnets. It is noteworthy that different hyperparameters result in different feature contributions and shape functions, and research with a larger dataset is essential to enhance robustness of results.

A third limitation relates to data normalization. To enable meaningful RI comparisons across datasets, we applied zero-mean and unit-variance normalization. Ideally, under consistent recording conditions, uniform speech task instructions, and robust feature extraction methods, such analyses could proceed without shifting the underlying data distribution.

A key limitation of this study is that AD and PD detection tasks were addressed separately due to differences in the publicly available datasets. These datasets vary in speech tasks and recording conditions, which substantially influence the features extracted~\cite{botelho2023ref_speech, botelho2022healthy_speech}. Future work could explore simultaneous detection using the proposed framework when datasets with comparable tasks and recording conditions become available.

Finally, it is important to to emphasize that the diseases, mechanisms, and their corresponding effects on the speech signal illustrated Fig.~\ref{fig:multi_diseases_impact_speech} serve as an initial foundation. We propose that further research and interdisciplinary dialogue are necessary to refine and expand this figure.

%%%%%%%%%%%%%%%%%%%%%%%%%%%%%%%%%%%%%%%%%%%%%%%%%%%%%%
\section{Conclusion}
\label{sec:conclusion}

This work introduced a framework for the use of speech as an interpretable biomarker for multiple diseases. Although focusing on Alzheimer's and Parkinson's diseases, the proposed framework is suitable for using speech as a biomarker in general, including the detection of other speech affecting diseases or even general health perturbations not typically categorized as diseases. This work started by discussing that speech affecting diseases should not be regarded individually for two reasons: (1)  they often have overlapping effects on the speech signal, and (2) they often are risk factors for each other. Therefore, it is argued here that a valuable first step is to characterize the speech of a reference population. This characterization was based on reference intervals, a concept common in clinical laboratory science, but novel in the field of speech analysis for disease detection. 
In this study, reference intervals were established for a reference population. Nevertheless, our vision encompasses the potential of individualized definition of reference speech. This self-definition would facilitate precise identification of early signs of disease, and would enable personalized healthcare.

The initial feature set was defined to capture manifestations of various speech-affecting diseases, focusing exclusively on interpretable features. 
However, a high-dimensional feature space complicates result interpretation. Thus, feature selection was based on a reference population, rather than the disease detection datasets, to establish a reference speech feature set and to avoid possible overfitting frequently observed with supervised feature selection on small datasets~\cite{dernoncourt2014analysis, soares2016feature, vabalas2019machine}.
Future research should expand the initial feature set to include additional knowledge-based features, capturing broader dimensions of reference speech, and further validate these features with other corpora for disease detection.

Finally, the definition of reference speech was leveraged for the detection of AD and PD, by comparing how much controls and patients deviate from the reference population.
Although the classification performance falls below other works in the literature, we advocate for the exploration of this approach due to its transparency, thereby advancing speech as a reliable biomarker.
In fact, it is well-documented that small sample sizes in clinical speech analysis studies often lead to overoptimistic estimates of model performance~\cite{berisha2022reported, ozbolt2022things}. Therefore, we underscore the importance of interpretable outcomes. 
Particularly, the shape-functions learnt by NAMs correspond exactly to the decision process, instead of \textit{a posterior} explanations. 
This transparency is crucial not only as a "second opinion" for clinicians, but also for early-stage research into speech as a biomarker. It facilitates multidisciplinary discussions among teams regarding the validity of model assumptions, and informs decisions regarding subsequent iterations, including data collection and feature refinement. Moreover, NAMs are suitable for multitask learning, enabling simultaneous detection of multiple diseases provided there is annotated speech data across different diseases, for the same speech tasks.

%%%%%%%%%%%%%%%%%%%%%%%%%%%%%%%%%%%%%%%%%%%%%%%%%%%%%%
\section{Acknowledgements}

This work was supported by Portuguese national funds through Fundação para a Ciência e a Tecnologia, with references DOI: 10.54499/UIDB/50021/2020 and SFRH/BD/149126/2019, and by the Portuguese Recovery and Resilience Plan and Next Generation EU European Funds, through project C644865762-00000008 (Accelerat.AI).

%%%%%%%%%%%%%%%%%%%%%%%%%%%%%%%%%%%%%%%%%%%%%%%%%%%%%%%%%%%%%%%%%%%%%%%%%%%%%%%%%%%%%%%%%%%%%%%%%%%%%%%%%%%%%%%%%%%%%%%%%%%%%%%%%%%%%%%%%%%%%%%%
\bibliographystyle{unsrt} 

\bibliography{./Bibliography}

\begin{IEEEbiography}
[{\includegraphics[width=1in,height=1.25in,clip,keepaspectratio]{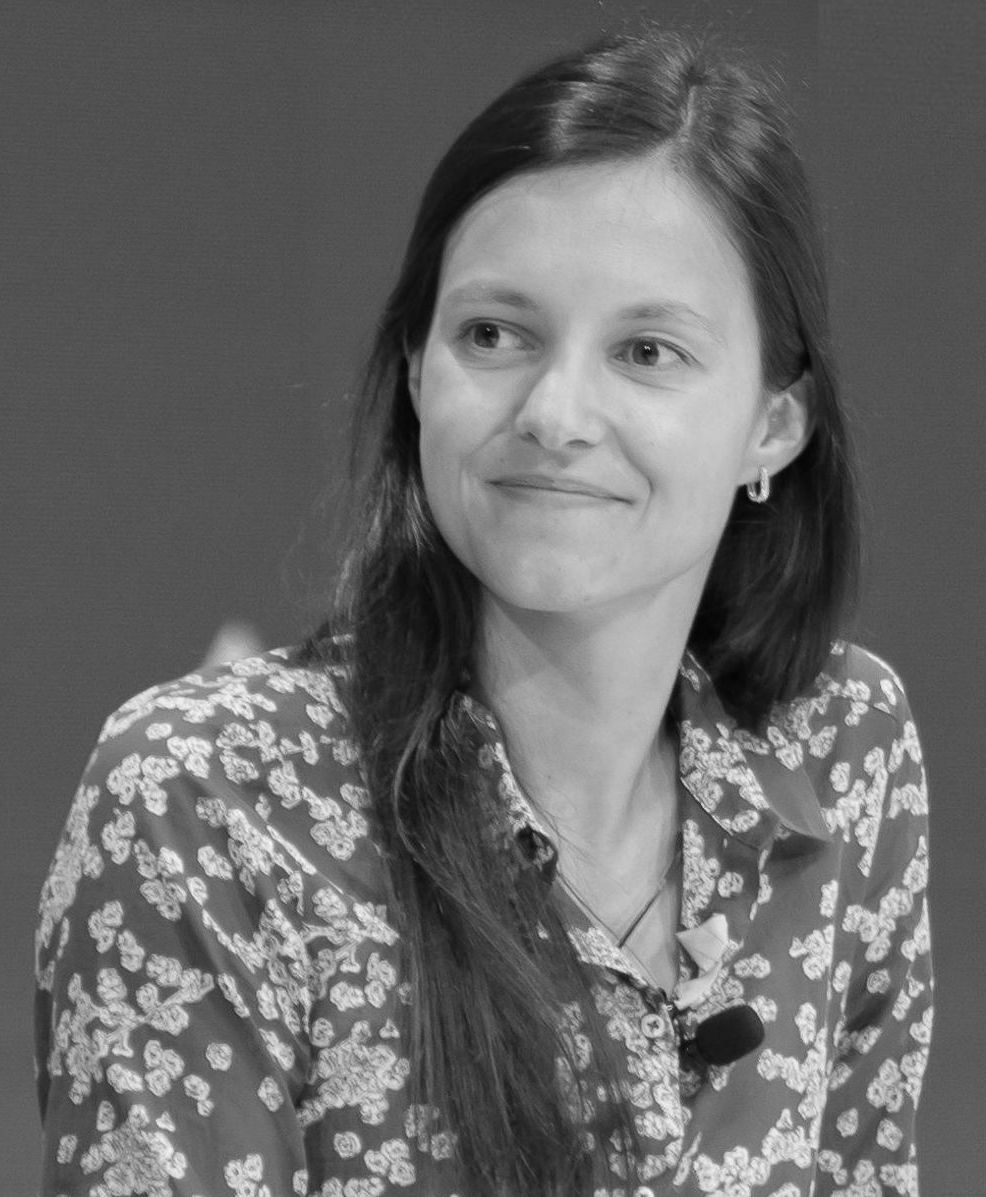}}]{Catarina Botelho} received the B.S. and M.S. degrees in Biomedical Engineering from Instituto Superior Técnico (IST), University of Lisbon, in 2018. Currently, she is a doctoral candidate in  Electrical and Computer Engineering, at INESC-ID and Instituto Superior Técnico. She was a research intern at Google AI, Toronto, and a visitor researcher at the Cognitive Systems Lab, University of Bremen. She was involved in the student advisory committee of the International Speech Communication Association (ISCA-SAC), from 2020 to 2023, acting as Coordinator in 2022.
Her scientific interests focus on speech and language technology and applied machine learning for healthcare.
\end{IEEEbiography}

\begin{IEEEbiography}[{\includegraphics[width=1in,height=1.25in,clip,keepaspectratio]{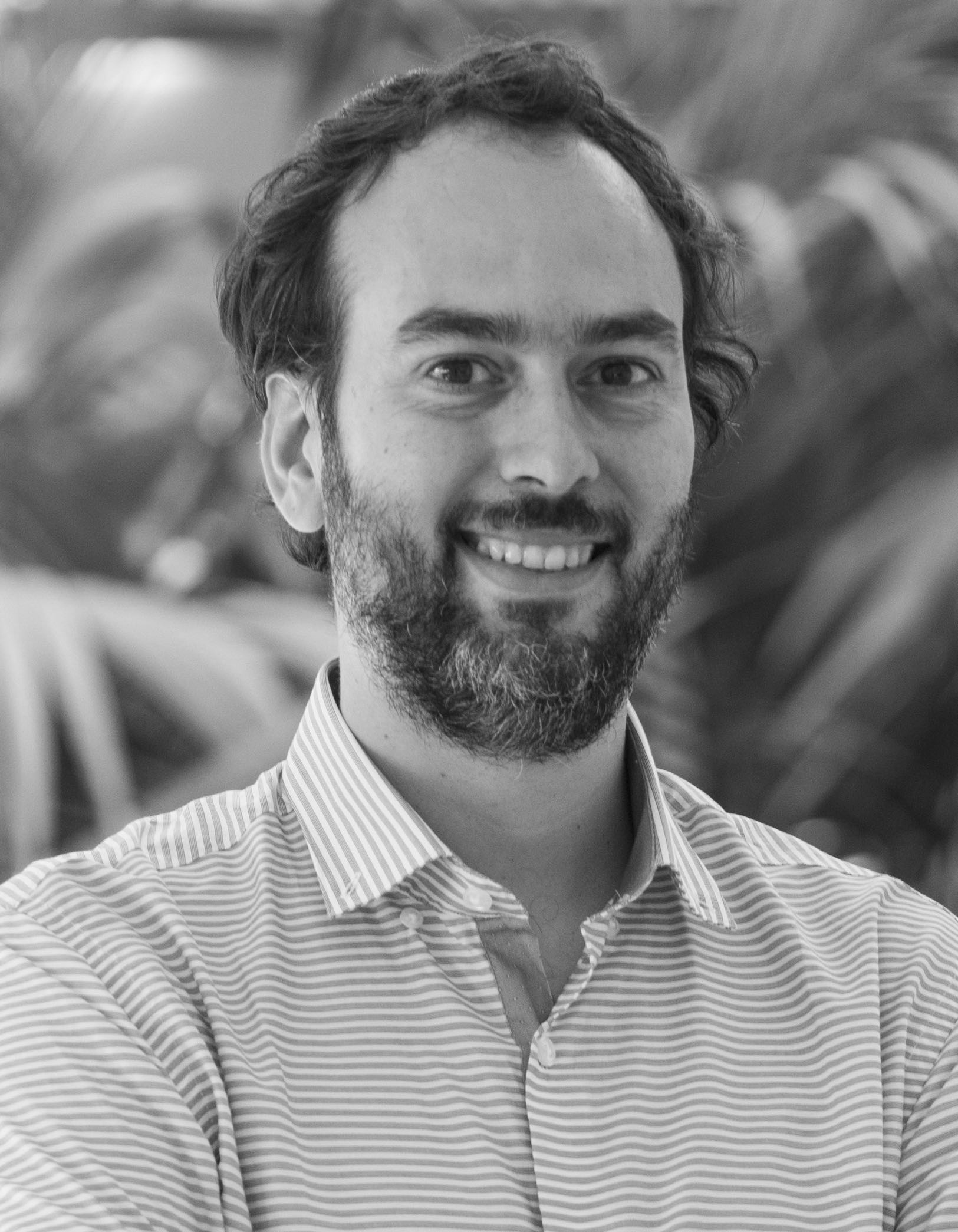}}]{Alberto Abad}
received the Telecommunication Engineering degree from the Technical University of Catalonia (UPC), Barcelona, Spain, in 2002 and the Ph.D. degree from UPC, in 2007. Currently, he is an Associate Professor at the Department of Computer Science and Engineering (DEI) of Instituto Superior Técnico (IST) and a researcher at INESC-ID. He is the coordinator of the Human Language Technologies laboratory at INESC-ID and the deputy coordinator of the Master in Computer Science and Engineering of IST. He is also an IEEE Senior member.
His research interests include robust speech recognition, speaker and language characterization, applied machine learning, healthcare applications, and privacy-preserving speech processing and machine learning.
\end{IEEEbiography}

\begin{IEEEbiography}[{\includegraphics[width=1in,height=1.25in,clip,keepaspectratio]{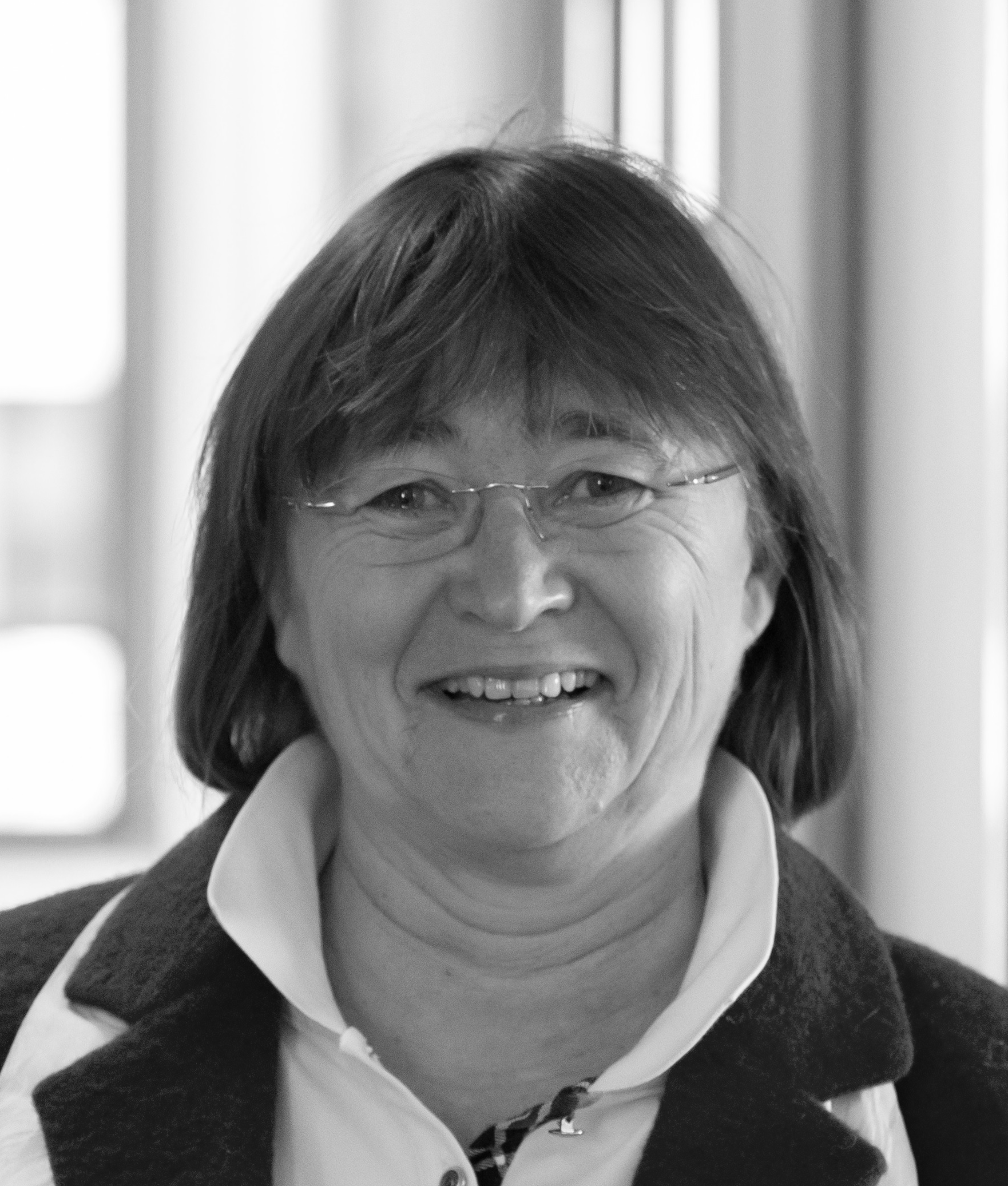}}]{Tanja Schultz}
received the diploma and doctoral degrees in Informatics from University of Karlsruhe, Germany, and spent over 20 years as Researcher and adjunct Research Professor at Carnegie Mellon University, PA USA. Since 2015 she is Professor for Cognitive Systems at the University of Bremen, Germany. In 2007, she founded the Cognitive Systems Lab where she and her team combine machine learning methods with innovations in biosignal processing to create biosignal-adaptive cognitive systems. She received several awards for her work and is a fellow of ISCA (2016), EASA (2017), IEEE (2020), and AAIA (2021). Currently, she leads the University's high-profile area “Minds, Media, Machines”, is a speaker of the DFG Research Unit Lifespan AI, and co-speaker of two research training groups. Recently, she established the international Master's program on Artificial Intelligence and Intelligent Systems. 
\end{IEEEbiography}

\begin{IEEEbiography}[{\includegraphics[width=1in,height=1.25in,clip,keepaspectratio]{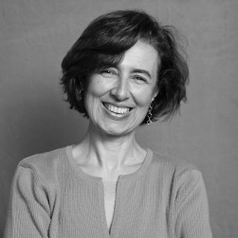}}]{Isabel Trancoso} is a former full professor at IST (University of Lisbon) and President of the Scientific Council of INESC-ID. She got her PhD in ECE from IST in 1987. She chaired the ECE Department of IST. She was Editor-in-Chief of the IEEE Transactions on Speech and Audio Processing and had many leadership roles in SPS (Signal Processing Society of IEEE) and ISCA (International Speech Communication Association), namely having been President of ISCA and Chair of the Fellow Evaluation Committees of both SPS and ISCA. Although recently retired, she is still actively supervising students and playing relevant roles in professional associations, such as Vice-Chair and Chair of the IEEE Fellow Committee (2023, 2024). She was elevated to IEEE Fellow in 2011, and to ISCA Fellow in 2014.
\end{IEEEbiography}

%%%%%%%%%%%%%%%%%%%%%%%%%%%%%%%%%%%%%%%%%%%%%%%%%%%%%%%%%%%%%%%%%%%%%%%%%%%%%%%%%%%%%%%%%%%%%%%%%%%%%%%%%%%%%%%%%%%%%%%%%%%%%%%%%%%%%%%%%%%%%%%%

\vfill
\pagebreak

\appendices

\section{Data pre-processing and Feature extraction}
\label{sec:appendix_data_preprocess_feature_ext}

\subsection{Sustained vowel pre-processing}

As described in \textit{step 1}, in section V.A, CLAC is a crowdsourced corpus, thus it includes recordings that exhibit anomalies, particularly in the recordings of sustained vowels. Examples of these anomalies include recordings with very little energy, recordings with a decrease in gain after a couple seconds because the tool used for data collection did not recognize the sound as speech, etc.
For this reason, and to improve the overall quality of the recordings that constitute the reference population, 
data filtering was performed to remove or segment  sustained vowel recordings from CLAC, following the twelve pre-processing steps described in Table~\ref{tab:vowel_preprocess}. The parameters (e.g. minimum acceptable RMSE of 0.005, the threshold of 15\% for abrupt changes, and the threshold of 100 Hz for standard deviation of F0) were  empirically defined. 
Steps 1-12 were applied to the sustained vowel recordings in CLAC, resulting in the exclusion of 355 (out of 2811)  recordings. Step 11 was applied to PC-GITA, resulting in the exclusion of 12 vowel recordings, because no "stable" segment was identified. In this dataset, there are no files with a standard deviation of F0 larger than 100 Hz.

\begin{table}[h]
  \caption{Sustained vowel pre-processing steps.}
  \vspace{-7pt}
  \label{tab:vowel_preprocess}
  \centering
  \setlength{\tabcolsep}{7pt}
  \resizebox{\linewidth}{!}{
  \begin{tabular}{p{0.1cm} p{4.8cm} p{4.5cm}}
    \toprule
    &  \textbf{Step} & \textbf{Notes}  \\
    \midrule
    1: & Exclude any files whose maximum RMSE is below 0.005. & \\
    2: & Search for abrupt changes in the RMSE. & Changes above 15\% of the RMSE of the signal are considered abrupt changes.  \\
    3: & \textbf{IF} no abrupt changes are detected  & \\
    4: & \hspace{0.1cm} Keep the recording. & \\
    5: & \textbf{ELSE}  & \\
    6: & \hspace{0.1cm} Extract the segments between each abrupt change. & \\
    7: & \hspace{0.1cm} Discard the segment after the last abrupt change until the end of the recording. & This segment may correspond to a gain reduction. \\
    8: & \hspace{0.1cm} Select the segment with the largest duration (without abrupt RMSE changes) for analysis. & \\
    9: & \textbf{ENDIF}  & \\
    10: & Split all segments longer than 4 seconds into chunks of 3~s, with a sliding window of 2~s. & To approximate the average duration of vowel segments in PC-GITA. \\
    11: & Before feature extraction, identify a "stable" sustained vowel segment. &  A "stable" segment corresponds to at least 110 periods without voice breaks. A voice break is considered if a period is larger than the maximum phonation period, set to 0.02~s. This criteria was defined based on a speech pathologist advice. \\
    12: & After feature extraction, exclude the segment if the standard deviation of F0 is larger than 100 Hz. & \\
    \bottomrule
  \end{tabular}
  }
  \vspace{-6pt}
\end{table}

%%%%%%%%%%%%%%%%%%%%%%%%%%%%%%%%%%%%%%%%%%%%%%%%%%%%%%%%%%%%%%%%%%%%%%%%%
\subsection{Audio samples excluded during data pre-processing and feature extraction}

Besides vowel segmentation, other steps in the method resulted in the exclusion of audio samples from the original pool of available samples.

The wav2vec ASR system failed to produce an output for 6 files in ADReSS (3 from the train set and 3 from the test set).

Additionally, for some data samples, it was not possible to extract all linguistic features. Particularly, for 9 wav2vec transcriptions in ADReSS (4 in the test set), 
15  wav2vec transcriptions in CLAC, and 
16  whisper transcriptions in  CLAC. Feature extraction failed because the generated transcriptions were either too short (for example, did not contain more than 14 tokens to compute coherence), or no English words were recognized.

The outlier removal method excluded 70 samples (28 vowels and 42 picture description) from CLAC, out of the pool of 4,992 audio samples for which it was possible to extract all features. 
These outliers represent approximately 1\% of the audio samples.

The excluded samples that belong to ADReSS and PC-GITA were not used to train the classifiers nor to define the best hyperparameters. 
However, to ensure a fair comparison with previous literature, these excluded samples are arbitrarily assigned the prediction "control" when reporting performance on the test set. This reflects a 50\% a priori probability of a correct prediction due to the balanced nature of the datasets.

%%%%%%%%%%%%%%%%%%%%%%%%%%%%%%%%%%%%%%%%%%%%%%%%%%%%%%%%%%%%%%%%%%%%%%%%%
\section{The feature "ratio of ambiguous coreference chains"}
\label{sec:appendix_ambiguous_coref_chain}

\begin{figure}[t]
  \centering
  \includegraphics[width=0.99\linewidth]{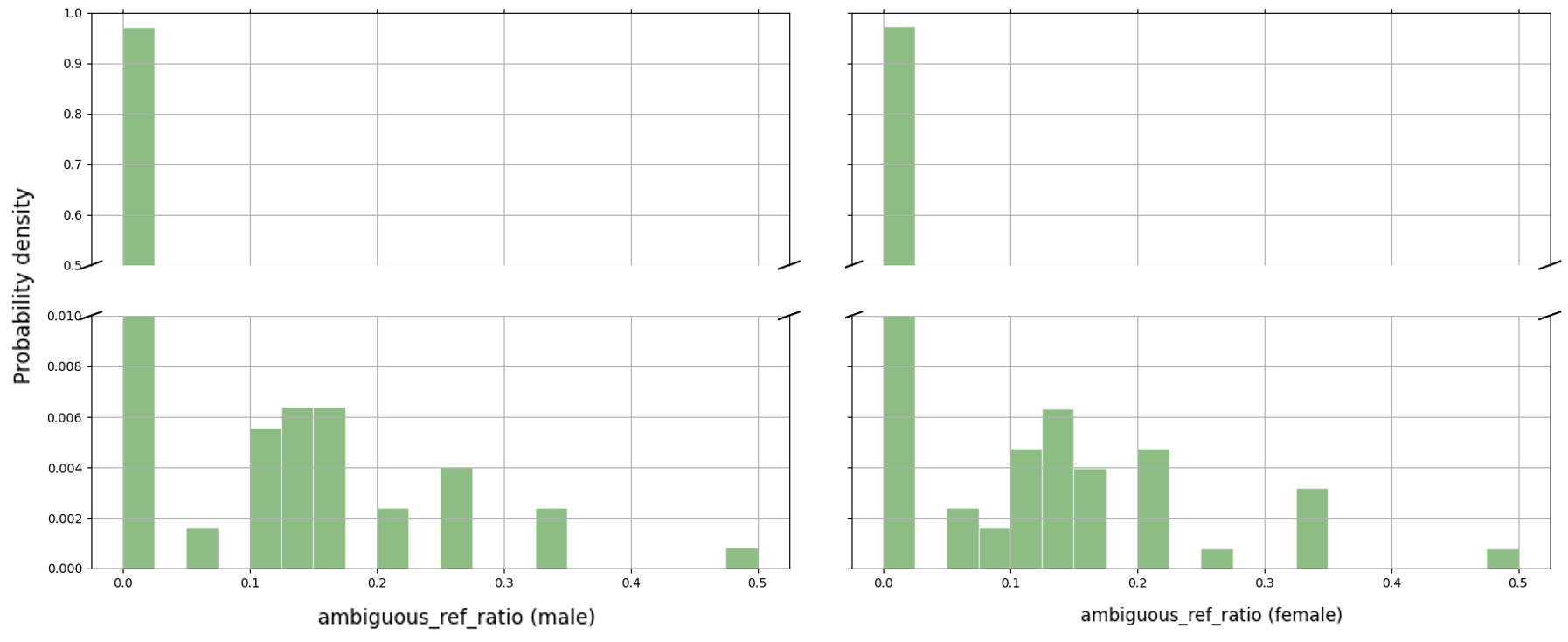}
  \caption{Distribution of \textit{ratio of ambiguous correference chains} on the reference population, based on whisper transcriptions.}
  \label{fig:invalid_ris_feats}
\end{figure}

As described in section V.B of the main document, the feature \textit{ratio of ambiguous coreference chains} was excluded from the analysis because the confidence intervals on the upper bound for both genders and both  ASR systems were larger than the RI itself, which indicates a poor confidence on the derived RI. 
Fig.~\ref{fig:invalid_ris_feats} shows the distribution of the \textit{ratio of ambiguous coreference chains} on the reference population. It is clear that the bulk of the distribution is very narrow as most samples correspond to zero. This resulted in a very narrow reference interval, and a very large confidence interval on the upper limit upon bootstrapping.
Intuitively, one can understand that in a healthy population describing an image, there would rarely be any ambiguous pronouns, i.e., entities not explicitly mentioned or mentioned only cataphorically. 
Table~\ref{tab:ambiguous_pronouns} shows examples of picture descriptions in CLAC and the corresponding coreference chains identified by the coreference resolver.

Although the feature was excluded from further analysis, it may be an interesting feature to explore, mainly when studying the speech of schizophrenia patients~\cite{bedi2015psychosis,iter2018schizophrenia}.

\begin{table}[t]
\footnotesize
\centering
\caption{Examples of picture descriptions in CLAC and the corresponding coreference chains identified by the coreference resolver. \label{tab:ambiguous_pronouns}}
\setlength{\tabcolsep}{4pt}
\resizebox{\linewidth}{!}{
\begin{tabular}{p{0.9cm} p{6cm} p{1cm}}
\toprule
\textbf{ASR} & \textbf{Description} & \textbf{Coref. chains} \\
\midrule
wav2vec & "i see a mother not paying attention to what's happening in her kitchen she's drying the dishes and appears to be day dreaming while looking out the window because she's not paying attention the sink is overflowing causing a flood in the kitchen and her children are about to steal cookies from a cabinet well most likely about to get hurt because the stoolis about to tip over." & ['Mother', 'Her', "She'S", "She'S", 'Her'], ['Kitchen', 'Kitchen'] \\

whisper & "A young boy is walking down a path while attempting to fly a kite and his dog is following him. Behind him, there is a lake with a young girl on the beach, building a sand castle. On that same lake there is a gentleman on a dock, landing a fish. And on that lake out in the distance, there's a sailboat sailing. Meanwhile, in the foreground, there is a couple having a picnic. The woman is pouring a glass of wine. There's a stereo playing, and the man is reading a book. Down the street, there is a house with a car in the driveway and a tree in the front yard and a flag at Polstaff." & ['Boy', 'His', 'Him', 'Him'], ['Lake', 'Lake', 'Lake'] \\

\bottomrule
\end{tabular}
}
\vspace{-8pt}
\end{table}

%%%%%%%%%%%%%%%%%%%%%%%%%%%%%%%%%%%%%%%%%%%%%%%%%%%%%%%%%%%%%%%%%%%%%%%%%%%%%%%%%%
\section{Quantitative comparison of reference and disease-affected speech}
\label{sec:appendix_ri_quantitative_analysis}

\begin{table}[t]
\centering
\caption{
(i) Number of audio samples. (ii) Average number of features outside of the RIs, per audio sample. (iii) Average feature distance to the limit of the RI, per audio sample. Because we could not assume the values follow a Normal distribution, $p-values$ 
were computed with a Mann-Whitney U test. (*) indicates statistically significant differences between controls and patients. $C$ 
and $P$ stand for controls and patients, and $F$ and $M$ stand for female and male speakers.
\label{tab:avg_feats_out_ri}}
\setlength{\tabcolsep}{5pt}
  \resizebox{0.99\linewidth}{!}{
\begin{tabular}{c cc|ccc|ccc}
\toprule
  & \multicolumn{2}{c|}{\textbf{(i) \#Samples}} & \multicolumn{3}{c|}{\textbf{(ii) \#Feats outside the RI}} & \multicolumn{3}{c}{\textbf{(iii) Feat dist to the RI}}  \\
 & C & P & C & P & \textit{p-value} & C & P & \textit{p-value} \\
\midrule
\multicolumn{4}{l}{\textbf{PC-GITA}} \\
\midrule
F & 75 & 74 & 1.5 & 3.5 & 0.0092* & 0.006 & 0.131 & 0.0003* \\
M & 70 & 69 & 1.4 & 2.0 & 0.327   & 0.008 & 0.038 & 0.0392* \\
 \midrule
\multicolumn{4}{l}{\textbf{ADReSS}} \\
\midrule
F & 43 & 43 & 3.0 & 4.8 & 0.0034* & 0.014 & 0.050 & 0.0001* \\
M & 35 & 35 & 3.3 & 4.6 & 0.004*  & 0.013 & 0.037 & 0.0005* \\
\bottomrule
\end{tabular}
}
\end{table}

\begin{figure*}[t]
  \centering
\includegraphics[width=0.8\linewidth]{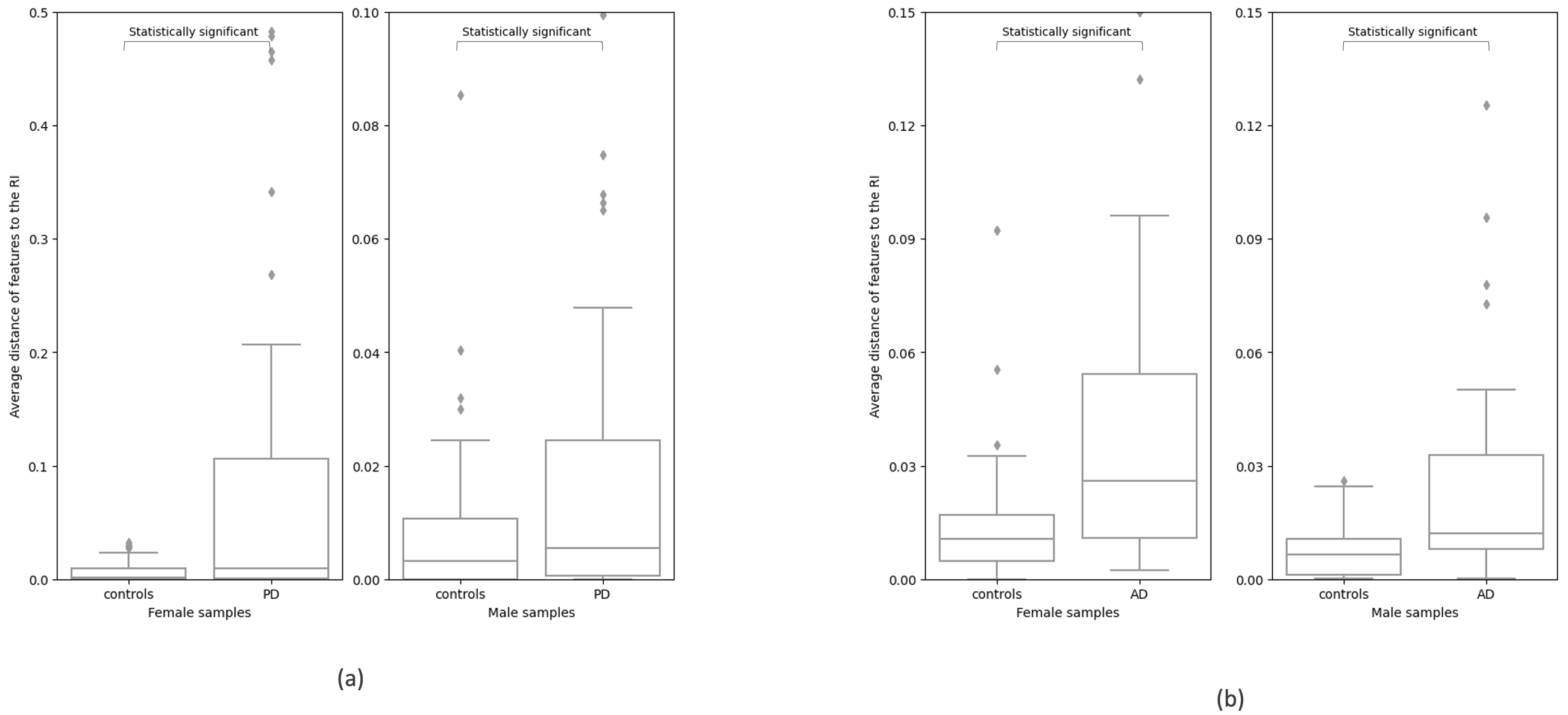}
  \caption{Distribution of the average distance of the features to the RI limits, per audio sample. (a) represents PC-GITA, and (b) represents ADReSS.}
  \label{fig:c6_feat_dist_boxplot}
\end{figure*}

The main document includes a qualitative comparison of reference speech and disease-affected speech. Here, a quantitative analysis is provided. 
Table~\ref{tab:avg_feats_out_ri} reports (i) the number of samples under analysis, (ii) the average number of features per audio sample that fall outside the RI, and (iii) the average distance of an audio sample's features from the RI limit.
The distance was computed as the difference between the feature value and the RI margin, divided by the length of the interval. If the feature value lies within the RI, the distance is considered 0.

The table shows that AD patients have a higher average number of features outside the RI per sample compared to control subjects. This difference is statistically significant in all cases, except for male speakers in PC-GITA.

More relevant than the number of features outside the reference interval, is the distance of these features from the interval. Table \ref{tab:avg_feats_out_ri} (iii) reports the average distance per group, and Fig.~\ref{fig:c6_feat_dist_boxplot} illustrates the distribution of these vales. This distance is, on average, higher for patients, and the difference between the two groups is considered statistically significant for all cases.

Future work could analyse this average distance per groups of features considered relevant for each disease. 

%%%%%%%%%%%%%%%%%%%%%%%%%%%%%%%%%%%%%%%%%%%%%%%%%%%%%%%%%%%%%%%%%%%%%%%%%%%%%%%%%%
\section{Deviation-scores}
\label{sec:appendix_deviation_scores}

The five deviation-scores ($DS$) compared were the following:
    
\begin{enumerate}
    \item $DS_{MSTD}$: 
    This deviation-score, inspired by~\cite{zusag23_carefulwhisper}, is based on the mean, $\mu$, and standard deviation, $\sigma$, of the feature distributions within the reference population. For each feature $i$, it is computed as $DS_{MSTD_{i}}=1-\frac{\sigma_{i}}{|\mu_{i}-x_{i}|}$ 
    if 
    $|\mu_{i}-x_{i}|>\sigma_{i}$ else it is set to 0.
    
    \item $DS_{MSTD-no-cap}$: This deviation-score is similar to the previous one, except it is not capped at 0 when the feature values are inside the interval 
    $[\mu_{i} - \sigma_{i}, \mu_{i} + \sigma_{i}]$. The idea with this deviation-score was to approximate it to the log-likelihood ratio.
    It is computed as: $DS_{MSTD-no-cap_{i}} = \frac{|\mu_{i} - x_{i}|}{\sigma_{i}}$.
    
    \item $DS_{Q123}$: This deviation-score is proposed as an alternative to $DS_{MSTD}$, that is based on the median ($Q_{2}$) and the first and third quartiles ($Q_{1}$ and $Q_{3}$), instead of the mean and standard deviation. 
    A second modification introduced in this deviation-score is that it yields negative or positive scores depending on whether the feature values fall below or above the interval $[Q_{1}, Q_{3}]$. This approach reflects the intuition that for certain features, deviating below or above the normal range does not carry the same implications, as discussed earlier in section~\ref{sec:c6-method-ref-int}--step 5. For each feature $i$, ${DS_{Q123_{i}}}$ is computed as:
    \begin{align} 
        DS_{Q123_{i}} = \begin{cases} 
        \frac{2|Q_{3i} - x_{i}|}{|Q_{3i} - Q_{1i}|} & \text{if } x_{i} > Q_{3i}, \\
        - \frac{2|Q_{1i} - x_{i}|}{|Q_{3i} - Q_{1i}|} & \text{if } x_{i} < Q_{1i}, \\
        0 & \text{elsewhere}.
        \end{cases}  
    \end{align}
    
    \item $DS_{RI}$: This deviation-score yields the same score for all feature values inside the reference interval, i.e., between the lower and upper bound of the reference interval, $[RI_{LB}, RI_{UB}]$. It also yields negative values for feature values below the reference interval. For each feature $i$, $DS_{RI_{i}}$ is computed as:
    \begin{align} 
        DS_{RI_{i}} = \begin{cases} 
        \frac{2|RI_{UBi} - x_{i}|}{|RI_{UBi} - RI_{LBi}|} & \text{if } x_{i} > RI_{UBi}, \\
        - \frac{2|RI_{LBi} - x_{i}|}{|RI_{UBi} - RI_{LBi}|} & \text{if } x_{i} < RI_{LBi}, \\
        0 & \text{elsewhere}.
        \end{cases}  
    \end{align}    
    
    \item $DS_{Mahalanobis}$: This deviation-score consists of computing the Mahalanobis distance of each new audio sample to the median ($Q_{2}$) of the reference population. Unlike the other deviation scores that are computed at the feature level, this deviation score is multivariate and considers the deviation of the vector of all features, \(x\) to the reference population. Thus, it provides a single value for each audio sample. It is computed as follows:
    $DS_{Mahalanobis} = \sqrt{(x-Q_{2})V^{-1}(x-Q_{2})^{T}}$, where $V$ is the covariance computed on the reference population. The Mahalanobis distance has been employed in~\cite{jiao_berisha2017interpretable} to compute the distance of a dysarthric speaker from the healthy distribution, based on phonological features.
\end{enumerate}

%%%%%%%%%%%%%%%%%%%%%%%%%
\section{Classification experiments with SVM and LR}
\label{sec:appendix_results_svm_lr}

Prior to the NAM experiments described in the main document, classification experiments were conducted using Support Vector Machines (SVMs) and Logistic Regression (LR). For these  experiments, two transcription types (whisper and wav2vec based), three normalization strategies (zero-mean and unit variance, \textit{MinMax}, and no normalization), six correlation threshold values, and five deviation-scores were compared. 
To evaluate whether the deviation scores provide an advantage over using directly the features as input to the classifiers, an extra scenario was also considered, where the features are directly fed to the classifiers after the stratified normalization strategy. 

The SVM hyperparameters were chosen based on a grid search on the development folds, although on a relatively small parameter space, to avoid getting high complexity models, which are more prone to overfitting. The hyperparameters compared were: $kernel \in {\text\{linear, rbf, poly\}}$, $C \in \{ 0.01, 0.1, 1\}$, and $degree\in {\text\{2, 3\}}$.

\begin{table*}[t]
  \caption{Best disease classification results, using SVM and logistic regression, in terms of accuracy (Acc), macro precision (P), macro recall (R), and macro F1, in [\%].}
  \vspace{-7pt}
  \label{tab:results_svm}
  \centering
  \resizebox{\linewidth}{!}{
  \begin{tabular}{ccccc cccccccccccc}
    \toprule
& & & & & \multicolumn{4}{c}{\textbf{Dev Folds}} & \multicolumn{4}{c}{\textbf{Test}} & \multicolumn{4}{c}{\textbf{Test -- Speaker MV}} \\
\textbf{CT} & \textbf{DT} & \textbf{norm} & \textbf{classifier} & \textbf{ASR} & \textbf{Acc} & \textbf{P} & \textbf{R} & \textbf{F1} & \textbf{Acc} & \textbf{P} & \textbf{R} & \textbf{F1} & \textbf{Acc} & \textbf{P} & \textbf{R} & \textbf{F1} \\
 \midrule
\multicolumn{2}{l}{\textbf{Parkinson's Disease}} \\
\midrule
1 & $DT_{RI}$ & MinMax & LR & -- & 72.2 &	72.7 &	72.2 & 	72.1 &	71.7 &	72.2 &	71.7 &	71.5 &	75.0 &	75.5 &	75.0 &	74.9 \\
0.9 & $DT_{Q123}$ & MinMax & SVM (linear, $C=0.01$) & -- & 69.4 &	69.9 &	69.4 &	69.2 &	69.7 &	70.2 &	69.7 &	69.5 &	77.0 &	77.9 &	77.0 &	76.8 \\
\midrule
\multicolumn{2}{l}{\textbf{Alzheimer's Disease}} \\
\midrule
1.0 & Raw features & MinMax & SVM (poly, $C=1$, $d=2$) & Whisper &  75.9 & 76.8 & 75.9 & 75.7 & 68.8 & 68.8 & 68.8 & 0.687 & -- & -- & -- & -- \\
0.7 & $DT_{RI}$ & None & LR & Whisper & 70.4 &	71.8 &	70.4 &	69.9 &	77.1 &	77.5 &	77.1 &	77.0 & -- & -- & -- & -- \\
\bottomrule
\end{tabular}
}
\vspace{-6pt}
\end{table*}

\begin{table}[t]
  \caption{Ablation study on the different variables for each configuration used in the disease detection experiments, using SVM and Logistic regression. Accuracy is reported in [\%]. $test^{\dag}$
  refers to the test set disregarding the files that were excluded during pre-processing or feature extraction.}
  \vspace{-7pt}
  \label{tab:results_svm_abaltion}
  \centering
  \resizebox{0.99\linewidth}{!}{
  \begin{tabular}{l cccccccc}
    \toprule

 & \multicolumn{4}{c}{\textbf{Parkinson's Disease}} &  & \multicolumn{3}{c}{\textbf{Alzheimer's Disease}} \\
 & \textbf{dev} & \textbf{test$^{\dag}$} & \textbf{test}  & \textbf{MV} &  & \textbf{dev} & \textbf{test$^{\dag}$} & \textbf{test} \\
 \midrule
 \multicolumn{4}{l}{\textbf{ASR}} \\
 \midrule
whisper & -- & -- & -- & -- &  & \textbf{62.0} & 63.7 & \textbf{63.7} \\
wav2vec & -- & -- & -- & -- &  & 58.5 & \textbf{65.4} & 60.0 \\

 \midrule
 \multicolumn{4}{l}{\textbf{DT}} \\
 \midrule
$DT_{MSTD}$ & 61.1 & 61.2 & 60.4 & 63.2 &  & 62.0 & 63.3 & 60.7 \\
$DT_{MSTD-no-cap}$ & 60.1 & 60.1 & 59.4 & 61.9 &  & 61.5 & 66.1 & 63.4 \\
$DT_{Q123}$ & \textbf{64.8} & \textbf{65.1} & \textbf{64.2} & \textbf{67.4} &  & 60.7 & 66.7 & 63.7 \\
$DT_{RI}$ & 60.8 & 61.2 & 60.4 & 62.6 &  & 60.7 & \textbf{66.9} & \textbf{64.1} \\
$DT_{Mahalanobis}$ & 60.2 & 59.4 & 58.7 & 61.5 &  & 54.1 & 58.8 & 56.6 \\
Raw features & 63.8 & 64.9 & 64.0 & 66.6 &  & \textbf{62.5} & 65.4 & 62.6 \\

 \midrule
 \multicolumn{4}{l}{\textbf{CT}} \\
 \midrule
$CT=0.5$ & \multirow{3}{*}{60.9} & \multirow{3}{*}{60.8} & \multirow{3}{*}{60.0} & \multirow{3}{*}{62.8} &  & \multirow{2}{*}{\textbf{60.9}} & \multirow{2}{*}{\textbf{66.1}} & \multirow{2}{*}{\textbf{63.2}} \\
$CT=0.6$ &  & & & &  & & & \\
$CT=0.7$ & & & & &  & 60.0 & 64.7 & 62.0 \\
$CT=0.8$ & 62.5 & 63.1 & 62.2 & 64.9 &  & 60.2 & 63.8 & 61.1 \\
$CT=0.9$ & 62.6 & 62.8 & 61.9 & 64.5 &  & 59.9 & 63.7 & 61.1 \\
$CT=1$ & \textbf{63.0} & \textbf{63.6} & \textbf{62.8} & \textbf{65.4} &  & 59.4 & 63.0 & 60.5 \\

 \midrule
 \multicolumn{4}{l}{\textbf{Normalization}}  \\
 \midrule
0-mean 1-var & 61.7 & 62.2 & 61.3 & 63.7 &  & 59.8 & 64.5 & 61.8 \\
MinMax & \textbf{65.5} & \textbf{65.5} & \textbf{64.5} & \textbf{68.2} &  & 59.5 & \textbf{65.6} & \textbf{62.7} \\
None & 58.2 & 58.3 & 57.6 & 59.7 &  & \textbf{61.4} & 63.5 & 61.0 \\

 \midrule
 \multicolumn{4}{l}{\textbf{Classifier}} \\
 \midrule
SVM & 61.6 & 61.3 & 60.5 & 63.1 &  & \textbf{61.9} & 64.2 & 61.5 \\
LR & \textbf{62.0} & \textbf{62.6} & \textbf{61.8} & \textbf{64.6} &  & 58.6 & \textbf{64.8} & \textbf{62.2} \\

\bottomrule
\end{tabular}
}
\end{table}

The best classification results on the development and test sets are reported in Table~\ref{tab:results_svm}. The results obtained on the complete set of configurations, including the different ASR systems, deviation-scores, correlation thresholds, and normalization strategies, are fully reported in~\cite{PhDthesis2024Botelho}. 

For PD detection, the best classification results on the development and test sets were achieved using deviation score $DS_{RI}$, without dimensionality reduction, with MinMax normalization, and a logistic regression classifier. This configuration achieved 71.7\% accuracy on the test set. 
If the 12 files excluded during data preprocessing (arbitrarily labeled as controls for a fair comparison with other works reporting on the entire dataset) were not included, the performance would increase to 73\%.
Each subject uttered 3 sustained vowels, thus performance was also evaluated at the speaker level, after computing the majority vote of the three predictions per subject, resulting in 75\% accuracy. The highest speaker-level accuracy (77\%) was achieved combining the deviation score $DT_{Q1,2,3}$, and the reduced dimensionality feature set with $CT=0.9$.

For AD detection in ADReSS, the best classification results on the development folds were obtained using directly the features that constitute the full dimensionality feature set ($CT=1$), combined with MinMax normalization, and an SVM classifier with second degree polynomial kernel, based on whisper transcriptions. This configuration reached 76\% accuracy on the development folds, and 69\% on the held-out test set.
The best results on the held-out test set were obtained using the deviation score $DT_{RI}$, and the reduced dimensionality feature set with $CT=0.7$. Pre-processing and feature extraction using whisper transcriptions did not lead to the exclusion of any files from analysis.
ADReSS only contains one picture description per subject, thus the performance reported at sample level is the same as at speaker level.

Given the numerous variables involved in these classification experiments, Table~\ref{tab:results_svm_abaltion} summarizes the average performance across all experiments for each ASR model, for each deviation-score, for each feature selection correlation threshold, for each normalization strategy, and for each classifier. 

Overall, whisper transcriptions yield better results than wav2vec transcriptions. This difference is partly due to wav2vec failing to generate a transcription for six files, and the extraction of linguistic features based on wav2vec transcriptions failing for nine additional files, which resulted in the exclusion of 15 files from further analysis. For consistency with other studies, the seven files in the test set were treated as if the prediction was control.

In terms of deviation-scores, the best average performance for PD detection was achieved by $DT_{Q1,2,3}$. This trend is not observed for AD. One partial explanation is that, with whisper transcriptions, it was not possible to compute the $DT_{Q1,2,3}$ for the \textit{discourse marker rate} feature. This issue arose because the bulk of discourse marker rate's data within the reference population corresponds to a very narrow range, leading to identical first and third quartiles, and resulting in an indeterminate $DT_{Q1,2,3}$. This feature is important for AD detection, as discussed in the main document, and its absence may hinder the performance of this deviation score.
Using the features directly as input to the classifier, i.e. without pre-computing a deviation-score, yielded the best average results on the ADReSS development folds. The best results on the ADReSS held-out test set were obtained with $DT_{RI}$.

Regarding the decision to reduce the dimensionality of the feature set based on Pearson correlation between features, it appears that the detection of PD benefits from using the entire feature set, while the detection of AD benefits from the reduced dimensionality feature set, with $CT=0.5$. Notably, the dimension of the entire feature set used to study sustained vowels (20 features) is very similar to the dimension of the reduced set used for studying the picture description task (23 features). 

The normalization strategy that appears to provide the best average results, for both PD and AD, is MinMax scaling, with the exception of the development folds in AD detection. Future work should further investigate the different normalization strategies, understand their impact, and discuss their inherent assumptions.

Finally, logistic regression appears to achieve better results than support vector machines. It is possible that the deviation-scores already provide substantial information, and thus a simpler classifier is sufficient. In fact, for AD detection, SVM achieves better results than LR on the development folds, but these are not generalized to the test set.

%%%%%%%%%%%%%%%%%%%%%%%%%%%%%%%%%%%%%%%%%%%%%%%%%%%%%%%%%%%%%%%%%%%%%%%%%
\section{Neural Additive  Models: architecture and hyperparameters}
\label{sec:appendix_nams_architecture}

The hyperparameter tuning for NAM was performed with Bayesian optimization using Gaussian Processes, as implemented in scikit-optimize~\cite{Head2021scikit-optmize}, with 100 calls to the optimizer. The hyperparameters considered for tuning, with minor variations from those described in \cite{agarwal2021nams}, were as follows: 

\begin{itemize}
    \item learning rate: $\{0.001, 0.002, 0.005, 0.01, 0.02, 0.05, 0.1\}$,
    \item dropout coefficient: $\{0, 0.05, 0.1, 0.2, 0.3, 0.4, 0.5, 0.6,$ $0.7, 0.8, 0.9\}$, 
    \item weight decay: $[0.000001, 0.0001]$, 
    \item feature dropout coefficient: $\{0, 0.05, 0.1, 0.2\}$
    \item output penalty coefficient: $[0.001, 0.1]$.
\end{itemize}

Additionally, the feature subnetworks were configured in one of the following ways: (i) one feedforward layer with 1024 hidden units, (ii) one feedforward layer with 512 hidden units, or (iii) three feedforward layers with 64, 64, and 32 hidden units.
The activation functions for the hidden units were either ReLU or ExU, as introduced by \cite{agarwal2021nams}. The batch size was set to 64.

Agarwal et al. \cite{agarwal2021nams} suggested using an ensemble of 10 to 100 models for each NAM. In this work, given the 10-fold cross-validation setting, we defined each NAM as an ensemble of 3 models, which after cross-validation results in a total of 30 models. Future work should investigate increasing the number of models.

Table~\ref{tab:nams_optimal_hp} reports the hyperparameters that yielded the best performance for the classification of PD and AD.

\begin{table}[t]
  \caption{Best parameters found for NAMs on classification of PD and AD, on PC-GITA and ADReSS, respectively. “Hidden units” shows the number of hidden layers as well as the number of neurons used in each layer for each feature network.}
  \vspace{-7pt}
  \label{tab:nams_optimal_hp}
  \centering
  \small
  \begin{tabular}{lcc}
    \toprule
 & PC-GITA & ADReSS \\
\midrule
Learning rate & 0.01 & 0.1 \\
Dropout & 0.8 & 0.6 \\
Weight decay & $1\times10^{-6}$ & $1\times10^{-6}$ \\
Feature dropout & 0.0 & 0.2 \\
Output penalty & 0.001 & 0.001 \\
Num units & 1024 & 1024 \\
Activation & ExU & ExU \\
\bottomrule
  \end{tabular}
\end{table}

\end{document}